\documentclass[11pt]{article}
\usepackage{amsmath,amsthm,epsfig,amsfonts,color,hhline, bm, graphicx,graphicx,fancyhdr,mathrsfs,subcaption}
\usepackage{rotating}
\allowdisplaybreaks
\newcommand{\VaR}{\mathrm{VaR}}
\newcommand{\ES}{\mathrm{ES}}

\newcommand{\p}{\mathbb{P}}
\newcommand{\E}{\mathbb{E}}
\DeclareMathOperator*{\argmin}{argmin}

\usepackage{comment}
\usepackage{booktabs}
\usepackage{setspace}
\usepackage{array}
\usepackage{amsmath}
\usepackage{amssymb}
\usepackage{url}
 \usepackage[colorlinks=true,citecolor=blue]{hyperref}
\usepackage{indentfirst}
\usepackage{enumerate}
\usepackage{comment}
\usepackage{dsfont}
\usepackage{natbib}
\usepackage[misc]{ifsym}
\usepackage[toc,page]{appendix}
\usepackage{multirow}
\usepackage{float}
\usepackage{todonotes}
\usepackage{hyperref}

\newtheorem{theorem}{{Theorem}}
\newtheorem{lemma}{{Lemma}}
\newtheorem{remark}{{Remark}}
\newtheorem{definition}{{Definition}}

\def\marginnote#1{\setbox0=\vtop{\hsize4pc
\small\raggedright\noindent\baselineskip9pt \rightskip=0.5pc plus
1.5pc #1}\leavevmode \vadjust{\dimen0=\dp0
\kern-\ht0\hbox{\kern-4.00pc\box0}\kern-\dimen0}}
\def\lboxit#1{\vbox{\hrule\hbox{\vrule\kern6pt
\vbox{\kern6pt#1\kern6pt}\kern6pt\vrule}\hrule}}

\usepackage{xr}
\makeatletter
\newcommand*{\addFileDependency}[1]{
  \typeout{(#1)}
  \@addtofilelist{#1}
  \IfFileExists{#1}{}{\typeout{No file #1.}}
}
\makeatother

\newcommand*{\myexternaldocument}[1]{%
    \externaldocument{#1}%
    \addFileDependency{#1.tex}%
    \addFileDependency{#1.aux}%
}
\myexternaldocument{Final_Supplementary}

\begin{document}
\thispagestyle{empty}
\title{A Revisit of the Optimal Excess-of-Loss Contract}

\author{Ernest Aboagye\thanks{J. Mack Robinson College of Business, Georgia State University, USA. Email addresses: \texttt{eaboagye1@gsu.edu} (Ernest Aboagye), \texttt{tfung@gsu.edu} (Tsz Chai Fung), \texttt{lpeng@gsu.edu} (Liang Peng), \texttt{qwang30@gsu.edu} (Qiuqi Wang).}~~~
Vali Asimit\thanks{Bayes Business School, City, University of London, UK. Email address: \texttt{asimit@city.ac.uk}}~~~
Tsz Chai Fung$^*$\thanks{Corresponding author.}~~~
Liang Peng$^*$~~~
Qiuqi Wang$^*$}

	\date{\today }
	\maketitle

 \bigskip
\begin{quote}
\begin{center}
\textbf{Abstract}
\end{center}

\smallskip
\noindent It is well-known that Excess-of-Loss reinsurance has more marketability than Stop-Loss reinsurance, though Stop-Loss reinsurance is the most prominent setting discussed in the optimal (re)insurance design literature. We point out that optimal reinsurance policy under Stop-Loss leads to a zero insolvency probability, which motivates our paper. We provide a remedy to this peculiar property of the optimal Stop-Loss reinsurance contract by investigating the optimal Excess-of-Loss reinsurance contract instead.  We also provide estimators for the optimal Excess-of-Loss and Stop-Loss contracts and investigate their statistical properties under many premium principle assumptions and various risk preferences, which according to our knowledge, have never been investigated in the literature. Simulated data and real-life data are used to illustrate our main theoretical findings. 

\smallskip
\noindent \textit{Keywords and phrases}: Risk analysis, Optimal Insurance, Nonparametric Estimation.
\end{quote}

\section{Introduction}\label{sec:setup}

\subsection{Literature Review}\label{sec:setup:lit_rev}
Risk transfer is an effective risk management exercise and consists of transferring liabilities from one or multiple risk holders (known as \emph{insurance buyer(s)}) to another or multiple insurance carriers (known as \emph{insurance seller(s)}).  Finding the optimal contact between (amongst) two (or more than two) parties has received a huge amount of attention in the literature of actuarial science. A simple Google Scholar search on April 23, 2024 with the keywords ``optimal insurance" and  ``risk transfer" resulted in 2,830,000 and 5,810,000, respectively research outputs. This is not surprising since the optimality of such risk management exercise goes beyond understanding insurance liabilities.  This paper aims to contribute to the problem of optimal insurance contract of insurance liabilities, which has a very specific trait that is not shared with other sector-specific liabilities (e.g., financial liabilities) in the sense that the insurance liabilities do not have a liquid market so that their value is market-based valuation. \emph{Cost-of-Capital} (CoC) approach is a practical methodology for evaluating insurance liabilities, which are based on the cost of meeting the local capital requirements to hold such liabilities in that territory. In other words, CoC is a regulatory-based methodology that is used within the insurance sector.  

The optimal risk transfer problem is often understood in the optimal insurance literature as how an insurer and reinsurer would share the aggregate liability between the two insurance players so that the risk position of the insurer is optimized; the optimization from the reinsurer's point of view is also possible.  One may view the problem from both the insurer's and reinsurer's point of view, a case in which the analysis becomes a Pareto optimal insurance contract problem that is a long-standing strand of research established in economic theory with ramifications in insurance and risk literature, but also in the wider operations research field; an insurance perspective could be found in Ruschendorf (2013) and references therein. Other equilibrium concepts are possible; for example, Boonen and Ghossoub (2023) investigates the Bowley equilibrium with risk sharing and optimal reinsurance formulations, and focus on the common traits of Bowley optimality and Pareto efficiency under fairly general preferences.  Bespoke conditions could be imposed on the optimal (re)insurance contract besides the usual absence of moral hazard; one interesting setting is the so called Vajda condition that is discussed in 
Boonen and Jiang (2022).

Depending on the risk preferences, the optimal reinsurance literature is quite rich; e.g., Cai et al. (2008) and Cai and Tan (2007)  consider \emph{Value-at-Risk} (VaR) and \emph{Expected Shortfall} (ES) buyer's preferences, while quantile risk and expectile preferences are investigated in Asimit, Badescu and Verdonck (2013), and Cai and Weng (2016), respectively; Balb\'as, Balb\'as and Heras (2009) investigates some general risk preferences. The optimal contract from the buyer's point of view in the presence of the seller's counterparty default risk is discussed in  Chen (2024), Chi and Tan (2021), Cai, Lemieux, and Liu (2014), Asimit, Badescu and Cheung (2013), and Bernard and Ludkovski (2012). Regulatory considerations are discussed for example in Asimit, Chi, and Hu (2015) and Bernard and Tian (2009). Robust formulations are investigated for example in Asimit, Hu and Xie (2019), Asimit~\textit{et al.} (2017), Balb\'as, Balb\'as and Heras (2011), Boonen and Jiang (2024) and Gollier (2014), while Cai, Li and Mao (2023) provides a theoretical perspective to robust decision-making when preferences are ordered by distorsion risk measures which are considered in our paper and many other papers in the optimal (re)insurance literature.  Non-standard settings are considered in the literature; e.g., B\"{a}uerle and Glauner (2018) investigates the optimal transfer in an insurance network from an economic point of view, while Asimit~\textit{et al.} (2016) studies Solvency II capital efficiency through risk transfers within an insurance group. 

The optimal insurance problem under expected utility settings is often defined without making any assumption regarding the seller's \emph{premium principle}. When risk preferences are ordered by risk measures, then premium principle assumptions are required. Kaluszka (2001) studies the mean-variance premium principle, Asimit, Badescu and Verdonck (2013) investigate quantile risk premium principles, and
Chi and Tan (2013) consider general premium principles, though many other papers rely on certain premium principle assumptions that are specific to the buyer's risk preferences.

\subsection{Background and Problem Definition}\label{sec:setup:bg}

Throughout this paper, the insurance field is represented by $(\Omega,\mathcal F,\p)$, an atomless probability space, endowed with  $L^0:=L^0(\Omega,\mathcal F,\p)$, the set of all non-negative real-valued random variables on this probability space. 
Let $L^q$, $q\in\left[0,\infty\right)$, be the set of random variables with finite $q^{\text{th}}$ moment, and $L^{\infty}$ be the set of bounded random variables.
A risk measure $\varphi$ is a function that maps an element of $L^0$ to a (extended) real number, i.e. $\varphi:L^0\to\overline{\Re}$.
We recall below some properties for a generic risk measure and generic random variable $Y$ -- with \emph{cumulative distribution function} (cdf) $F_Y$, \emph{survival distribution function} $\bar{F}_Y$, and \emph{generalized left-continuous inverse} $F_Y^-(s):=\inf_{x\in\Re}\big\{F_Y(x)\ge s\big\}$ -- representing the future loss of a financial asset or insurance liability.  
\vspace{-0.25cm}
\begin{align*}
&\text{\it Convexity: } \varphi(a Y_1+(1-a)Y_2)\leq a\varphi(Y_1)+(1\!-\!a)\varphi(Y_2)\;\text{for any}\;Y_1,Y_2\in L^0 \;\text{and}\; a\in[0,1];\\
&\text{\it Homogeneous of order $\tau>0$: } \varphi\left(c Y\right)=c^{\tau}\varphi(Y) \;\text{for any}\; Y\in L^0 \;\text{and}\; c\ge0;\\
&\text{\it Shift invariance: } \varphi(Y+c)=\varphi(Y) \;\text{for any}\; Y\in L^0 \;\text{and}\; c\in\Re;\\
&\text{\it Translation invariance: } \varphi\left(Y+c\right)=\varphi(Y)+c \;\text{for any}\; Y\in L^0 \;\text{and}\; c\in\Re.
\end{align*}
These properties are well-known in the literature and an extensive introduction to risk measures can be found in F\"{o}llmer and Schied (2011). Two well-known risk measures are  \emph{Value-at-Risk} (VaR) and \emph{Expected Shortfall} (ES), defined as
\[\VaR_p(Y)=F_Y^-(p)~~\text{and}~~ \ES_p(Y)=\min_{t\in\Re} \left\{t+\frac{1}{1-p}\E(Y-t)_+\right\},\]
where $(\cdot)_+=\max(\cdot,0)$ and $p\in(0, 1)$ is the risk level.  
It is evident that the two risk measures are homogeneous of order $1$ and translation invariant, and  ES is convex. 
 
We are now ready to provide the mathematical formulation of the problem of interest.  Suppose that an insurer has insured a large number of policies with independent and identically distributed non-negative losses $X_i$ for $1\le i\le N$ with cdf $F_{X_1}(x)$. 

We consider now that the reinsurance premium is calculated by the expected value principle.  Thus, the total cost for this portfolio of policies after buying \emph{Excess-of-Loss (EoL)} reinsurance becomes
\begin{eqnarray}\label{s_3_def}
    T(d,N,\rho)=\sum_{i=1}^N (X_i\wedge d) +(1+\rho)\E\left(\sum_{i=1}^N (X_i-d)_+\right),
\end{eqnarray}
where $\rho>0$ is the loading factor. A practical question is to find the optimal retention $d$ for  $T(d,N,\rho)$ by minimizing the buyer's risk when its perception of risk is modeled by some given risk measures such as VaR and ES.  To better appreciate our study, we first point out an issue with the \emph{Stop-loss (SL)} optimal reinsurance (SL is EoL with $N=1$) in Cai and Tan (2007), where the total cost $T(d,1,\rho)$ is studied;  that is, the optimal retention is found via minimizing $\VaR_p\big(T(d,1,\rho)\big)$ or $\ES_p\big(T(d,1,\rho)\big)$, which leads to the following optimal retention
\begin{equation} \label{eq:opt_d_SL}
d^*=F_{X_1}^{-}\left(1-\frac 1{1+\rho}\right)~ \text{when}~1-p<(1+\rho)^{-1}.
\end{equation}
Now, if $1-p<(1+\rho)^{-1}$ then
\[\p\Big(T(d^*,1,\rho)>\VaR_p\big(T(d^*,1,\rho)\big)\Big)=\p(X_1\wedge d^*>d^*)=0,\]
implying no high risk to the buyer, which is mathematically explained by the truncated buyer's liability $X_1\wedge d$. The same issue remains if one replaces $X_1\wedge d$ by $(\sum_{i=1}^NX_i)\wedge d$, i.e., considering the SL for the total loss instead of one loss in Cai and Tan (2007).
However, when the number of policies is large enough, $\sum_{i=1}^N(X_i\wedge d)$ will not have such a truncation issue to cause a severely distorted risk level for optimal retention, and thus, the optimal EoL (with $N>1$) retention would not share the same counter-intuitive property as SL (when $N=1$). Specifically, under the EoL approach, $\p\left(T(d^*,N,\rho)>\VaR_p(T(d^*,N,\rho))\right)$ becomes $1-p$, the ``correct'' level, as long as $N$ is sufficiently large. Numerical and theoretical justifications of this assertion are leveraged to Section~\ref{supp:sec:num} of the supplementary material. Furthermore, the SL optimal retention $d^*$ in \eqref{eq:opt_d_SL} is not an explicit function of the risk level $p$. This is also counter-intuitive as the SL optimal retention may remain constant while $p$, which implies the insurance company's level of risk aversion, increases. Conversely, we will show in the real data analysis (see Section~\ref{sec:data}) that the EoL optimal retention decreases as $p$ increases.

To derive the optimal retention, it is necessary to know the distribution function of $\sum_{i=1}^N(X_i\wedge d)$, which is well-known to be challenging. This motivates us to define approximately optimal retention by using a normal distribution to approximate the distribution of $\sum_{i=1}^N(X_i\wedge d)$ when $N$ is large enough. 

Before outlining our main contributions, we would like to differentiate the EoL and SL contracts, which are compared in this paper. Note that EoL has more marketability than SL as the latter is prohibitively expensive to buyers since the deductible is applied to the annual aggregate loss and not to the individual claims (as for EoL).  There are other negative traits of SL that are not shared with EoL.  For example, the loss development of an insurance claim is the process of a claim from reporting until the claim is fully settled, which takes a significant amount of time for many lines of business such as personal accident insurance, medical malpractice insurance, workers compensation, liability claims, etc.; the lag is even larger for long-tail lines of coverage where arbitrage or court proceedings are more likely to occur.  Long lags are big impediments to activate SL contracts since the deductible is applied to the aggregate loss, which is known when all claims from that year are fully settled and that may require many years;  this is not the case to EoL where each claim is shared between the buyer and seller.

The main contributions of this paper are twofold. \emph{First}, we point out that optimal reinsurance policy under SL -- one of the most prominent settings discussed in the literature -- leads to zero insolvency probability for VaR-based regulatory environments as is the case for EU and UK insurance companies where capital requirements are designed on the 1/200 event basis over a one-year time horizon.  This peculiar property of the optimal SL reinsurance contract is the main motivation of our paper and we show that a remedy is possible if one investigates the optimal EoL reinsurance contract instead.  \emph{Second}, we provide estimators to the optimal EoL/SL deductible and investigate their statistical properties under many premium principle assumptions and various risk preferences, which, according to our knowledge, have never been investigated in the literature. 

The paper is organized as follows: EoL risk model is considered under the $\VaR_p$ risk measure in Section~\ref{sec:var} for various premium principles, which are further generalized in Section~\ref{sec:es} when the risk preferences are ordered by distortion risk measures. Some simulation studies are provided in Section~\ref{sec_sim}, while real data analysis is employed in Section~\ref{sec:data}.  

\section{Approximately optimal retention for VaR}\label{sec:var}

In this section, we consider the total cost of $T(d,N,\rho)$ under the $\VaR_p$ risk measure. Later, we will generalize the result to distortion risk measures in Section~\ref{sec:es}.  Because VaR is translation invariant, we have 
\begin{eqnarray}\label{T_2_EVP}
    \VaR_p\big(T(d,N,\rho)\big)=\VaR_p\left(\sum_{i=1}^N (X_i\wedge d)\right) + (1+\rho)N\E\{(X_1-d)_+\}.
\end{eqnarray}
Define
\[\left\{\begin{array}{ll}
\mu_1(d)=\E(X_1\wedge d)=\int_0^d\bar F_{X_1}(x)\,dx,\\
\mu_2(d)=\E(X_1^2\wedge d^2)=2\int_0^d\bar F_{X_1}(x)x\,dx,\\
\nu_1(d)=\E\{(X_1-d)_+\}=\int_d^{\infty}\bar F_{X_1}(x)\,dx.\end{array}\right.\]
For large $N$, it follows from the Central Limit Theorem that
\begin{equation}\label{appr}
\VaR_p\left(\sum_{i=1}^N (X_i\wedge d)\right)=N\mu_1(d)+\sqrt N\sqrt{\mu_2(d)-\mu_1^2(d)}\Phi^-(p)+o(\sqrt N),
\end{equation}
where $\Phi(x)$ is the cdf of a standard normal random variable.
Therefore, instead of minimizing $\VaR_p\big(T(d,N,\rho)\big)$ to obtain the optimal retention $d$, we propose to ignore the $o(\sqrt N)$ term in \eqref{appr} to approximate the right hand side of \eqref{T_2_EVP}, i.e.,  minimizing 
\begin{align}
    &G_{N,\rho}(d):=N\E(X_1)+N\rho\nu_1(d)+\sqrt{N}\sqrt{\mu_2(d)-\mu^2_1(d)}\Phi^-(p),\label{eq:EVP_prob2}
\end{align}
whose solution is called an \emph{approximately optimal retention}.

The structure of this section is as follows: Section~\ref{sec_2_1} identifies the optimal retention by assuming a constant loading factor under the expected value principle, while a decreasing loading factor is considered in Section~\ref{sec_2_2}; the standard deviation and Sharpe ratio principles are explored in Sections~\ref{sec_2_3} and \ref{sec_2_4} by assuming a decreasing loading factor.

\subsection{Constant Loading Factor}\label{sec_2_1}

We now solve \eqref{eq:EVP_prob2} with a constant loading factor $\rho>0$. In this case, the optimal retention turns out to be a solution to
\begin{equation}\label{H}
   H_{N,\rho}(d):=\{d-\mu_1(d)\}^2-\left(\frac{\sqrt N\rho}{\Phi^-(p)}\right)^2\{\mu_2(d)-\mu_1^2(d)\}=0.\end{equation} 
The next result stated as Theorem~\ref{thm:1} shows that \eqref{H} admits a unique solution under some very mild regularity conditions.  Recall that we allow $F_{X_1}(0)>0$ in Theorem~\ref{thm:1}, which means that the event of having no claim is not a null set.

\begin{theorem}\label{thm:1} 
Assume $\E(X_1)<\infty$, $F_{X_1}(\cdot)$ has the support $[0, \infty)$ (i.e., $F_{X_1}(0)>0$) or $(0, \infty)$, and is continuous on $(0, \infty)$. When the support is $[0, \infty)$, we further assume 
$F_{X_1}(0)<\frac{N\rho^2}{N\rho^2+(\Phi^-(p))^2}$, which is always true when $N$ is large enough. Then, there exists a unique approximately optimal retention $d^*_{N,\rho}\in(0, \infty)$ such that 
\[d^*_{N,\rho}=\argmin_{d>0}G_{N,\rho}(d)~~\text{and}~~H_{N,\rho}(d_{N,\rho}^*)=0.\]
{}
\end{theorem}
\begin{proof}
Note that $\mu_1'(d)=\bar F_{X_1}(d)$, $\mu'_2(d)=2d\bar F_{X_1}(d)$, and $\nu_1'(d)=-\bar F_{X_1}(d)$, and in turn,
\begin{eqnarray}\label{eq_G_H}
  G'_{N,\rho}(d)&=&-N\rho\bar F_{X_1}(d)+\sqrt{N}\bar F_{X_1}(d)\frac{d-\mu_1(d)}{\sqrt{\mu_2(d)-\mu_1^2(d)}}\Phi^{-}(p)\\
        &=&\frac{\sqrt N\bar F_{X_1}(d)\Phi^-(p)}{\left(\frac{d-\mu_1(d)}{\sqrt{\mu_2(d)-\mu_1^2(d)}}+\frac{\sqrt N\rho}{\Phi^-(p)}\right)\{\mu_2(d)-\mu_1^2(d)\}}H_{N,\rho}(d). \nonumber
\end{eqnarray}
Hence, solving $G'_{N,\rho}(d)=0$ for $d\in(0, \infty)$ is equivalent to solving $H_{N,\rho}(d)=0$ for $d\in(0, \infty)$. That is, we only need to show that there is a unique solution for $H_{N,\rho}(d)=0$.
     
Since $d-\mu_1(d)> 0$ and $\mu_2(d)-\mu_1^2(d)> 0$ for all $d> 0$, and 
   \[H'_{N,\rho}(d)=2\{d-\mu_1(d)\}\left(F_{X_1}(d)\left(1+\left(\frac{\sqrt N\rho}{\Phi^-(p)}\right)^2\right)-\left(\frac{\sqrt N\rho}{\Phi^-(p)}\right)^2\right),\]
we conclude that 
  $H'_{N,\rho}(d)$ is negative, zero, and positive for $0<d<d_1$, $d_1\le d\le d_2$, and
  $d>d_2$, respectively, where 
  $F_{X_1}(d)=\frac{N\rho^2}{N\rho^2+(\Phi^-(p))^2}$ happens and only happens on $d\in [d_1, d_2]$, which is ensured by the conditions that the right endpoint of $F_{X_1}(x)$ is infinity, $F_{X_1}(x)$ is continuous on $(0, \infty)$, and $F_{X_1}(0)<\frac{N\rho^2}{N\rho^2+(\Phi^-(p))^2}$. That is,
 \begin{equation}\label{pfTh1-2}
  H_{N,\rho}(d)~\text{is strictly $\downarrow$ on}~ (0, d_1),~\text{constant on}~ [d_1, d_2],~\text{and strictly $\uparrow$ on}~(d_2, \infty),
  \end{equation}
Note that
\[
  \lim_{d\to \infty}\frac{d^2}{\mu_2(d)}=\left\{
  \begin{array}{ll}
  \lim_{d\to\infty}\frac{2d}{2d\bar F_{X_1}(d)}=\infty &\text{if}~\mu_2(\infty)=\infty,\\
  \infty &\text{if}~\mu_2(\infty)<\infty.
  \end{array}\right.\]
Thus, $\lim_{d\to\infty}H_{N,\rho}(d)/d^2=1$ and 
$\lim_{d\to\infty}H_{N,\rho}(d)=\infty$. The latter, \eqref{eq_G_H} and \eqref{pfTh1-2}, and the fact that $\lim_{d\to0}H_{N,\rho}(d)=0$ conclude that $H_{N,\rho}(d)=0$  has a unique solution on $(d_2, \infty)$.  The proof is now complete.
\end{proof}
 
\subsection{Decreasing Loading Factor}\label{sec_2_2}
 
Note that $d_{N,\rho}^*$ diverges to infinity as $N\to\infty$, which is not surprising since a constant $\rho$ for any $N$ implies that the seller does not include the diversification effect in its premium calculation, case in which the seller would not be incentivized to participate in such reinsurance contract. Therefore, it would be more practical to adjust/reduce the loading factor as $N$ gets large so that the reinsurance premium becomes more realistic.  To estimate $d_{N,\rho}^*$ and study its asymptotic properties, we consider instead a bounded approximated optimal retention by assuming $\rho=\rho_N$ such that
\begin{equation}\label{con1}
 \lim_{N\to\infty}\rho_N\sqrt N=\delta\in(0, \infty),
\end{equation}
where $\delta$ may depend on the retention $d$. 
In this case, $d_{N,\rho_N}^*$ is the unique solution to
$H_{N,\rho_N}(d)=0$ and converges to the unique solution to
\[\{d-\hat\mu_1(d)\}^2-\left(\frac{\delta}{\Phi^-(p)}\right)^2\{\hat\mu_2(d)-\hat\mu_1^2(d)\}=0.\]
To estimate this solution nonparametrically, we solve the following equation
\begin{equation}\label{est}
\hat H_{N,\rho_N}(d):=\{d-\hat\mu_1(d)\}^2-\left(\frac{\sqrt N\rho_N}{\Phi^-(p)}\right)^2\{\hat\mu_2(d)-\hat\mu_1^2(d)\}=0,
\end{equation}
where
\begin{equation} \label{eq:mu_hat}
\hat\mu_1(d)=\frac 1N\sum_{i=1}^N (X_i\wedge d)~\text{and}~\hat\mu_2(d)=\frac 1N\sum_{i=1}^N(X_i^2\wedge d^2).
\end{equation}
Let $\hat d_{N,\rho_N}^*$ denote this solution, which is an estimator for $d_{N,\rho_N}^*$. Let $\Sigma(d)$ denote the covariance matrix of $\boldsymbol{Z}_i(d)$, where $\boldsymbol{Z}_i(d)=(X_i\wedge d, X_i^2\wedge d^2)^{\tau}$, and define
\begin{equation} \label{eq:mu_d_hat}
\hat\mu_1^*(d)=\frac 1N\sum_{i=1}^NI(X_i>d)~\text{and}~\hat\mu_2^*(d)=\frac{2d}N\sum_{i=1}^NI(X_i>d),
\end{equation}
which estimate the first-order derivatives, $\mu_1'(d)$ and $\mu_2'(d)$, respectively.  The asymptotic properties of $\hat d^*_{N,\rho_N}$ are given in Theorem~\ref{thm:2}.
  
\begin{theorem}\label{thm:2}
Under conditions of Theorem~\ref{thm:1} and \eqref{con1}, we have
\[\frac{\sqrt N\{\hat d^*_{N,\rho_N}-d^*_{N,\rho_N}\}}{\hat c_0^{-1}\sqrt{(\hat c_1, \hat c_2)\hat\Sigma_0(\hat c_1, \hat c_2)^{\tau}}}\overset{d}{\to} N(0, 1),\]
where 
\begin{eqnarray*}
&&\hat c_0=2\{\hat d^*_{N,\rho_N}-\hat\mu_1(\hat d^*_{N,\rho_N})\}\{1-\hat\mu_1^*(\hat d^*_{N,\rho_N})\}\\
&&\hspace{1cm}-\left(\frac{\rho_N\sqrt N}{\Phi^-(p)}\right)^2\{\hat\mu_2^*(\hat d^*_{N,\rho_N})-2\hat\mu_1(\hat d^*_{N,\rho_N})\hat\mu_1^*(\hat d^*_{N,\rho_N})\},\\
 &&\hat c_1=2\{\hat d^*_{N,\rho_N}-\hat\mu_1(\hat d^*_{N,\rho_N})\}-\left(\frac{\rho_N\sqrt N}{\Phi^-(p)}\right)^22\hat\mu_1(\hat d^*_{N,\rho_N}),~\hat c_2=\left(\frac{\rho_N\sqrt N}{\Phi^-(p)}\right)^2,\\
 &&\hat\Sigma_0=\frac 1N\sum_{i=1}^N\left[\boldsymbol{Z}_i(\hat d^*_{N,\rho_N})-\frac{1}{N}\sum_{i'=1}^N\boldsymbol{Z}_{i'}(\hat d^*_{N,\rho_N})\right]\left[\boldsymbol{Z}_i(\hat d^*_{N,\rho_N})-\frac{1}{N}\sum_{i'=1}^N\boldsymbol{Z}_{i'}(\hat d^*_{N,\rho_N})\right]^{\tau}.
 \end{eqnarray*}
 \end{theorem}
\begin{proof}
For simplicity, the proof uses $d^*$ and $\hat d^*$ for $d_{N,\rho_N}^*$ and $\hat d_{N,\rho_N}^*$, respectively.  Then
 \begin{equation}\label{normal1}
 \sqrt N\begin{pmatrix}\hat\mu_1(d^*)-\mu_1(d^*)\\
 \hat\mu_2(d^*)-\mu_2(d^*)
 \end{pmatrix}\overset{\mathrm{d}}{\to} N(\boldsymbol{0},\Sigma_0)
 \end{equation}
 when
 \begin{equation}\label{con2}
 \Sigma(d^*)\to\Sigma_0~\text{as}~N\to\infty.
 \end{equation}
 It follows from \eqref{normal1} that
 \[\begin{array}{ll}
 \hat\mu_1(\hat d^*)-\mu_1(d^*)&=\hat\mu_1(\hat d^*)-\mu_1(\hat d^*)+\mu_1(\hat d^*)-\mu_1(d^*)\\
 &=\{\hat\mu_1(d^*)-\mu_1(d^*)\}+\mu_1'(d^*)\{\hat d^*-d^*\}+o_p(1/\sqrt N),
  \end{array}\]
 \[
 \hat\mu_2(\hat d^*)-\mu_2(d^*)
 =\{\hat\mu_2(d^*)-\mu_2(d^*)\}+\mu_2'(d^*)\{\hat d^*-d^*\}+o_p(1/\sqrt N),
 \] 
 \[\begin{array}{ll}
 &\{\hat d^*-\hat\mu_1(\hat d^*)\}^2-\{d^*-\mu_1(d^*)\}^2\\
 =&2\{d^*-\mu_1(d^*)\}\{1-\mu_1'(d^*)\}(\hat d^*-d^*)\\
 &-2\{d^*-\mu_1(d^*)\}\{\hat\mu_1(d^*)-\mu_1(d^*))\}+o_p(1/\sqrt N),
 \end{array}\]
 \[\begin{array}{ll}
 &\{\hat\mu_2(\hat d^*)-\hat\mu_1^2(\hat d^*)\}-\{\mu_2(d^*)-\mu_1^2(d^*)\}\\
 =&\{\hat\mu_2(d^*)-\mu_2(d^*)\}-2\mu_1(d^*)\{\hat\mu_1(d^*)-\mu_1(d^*)\}\\
 &+\{\mu_2'(d^*)-2\mu_1(d^*)\mu_1'(d^*)\}\{\hat d^*-d^*\}+o_p(1/\sqrt N),
 \end{array}\]
 implying that
 \[\begin{array}{ll}
 0&=\hat{H}_{N,\rho_N}(\hat d^*)-H_{N,\rho_N}(d^*)\\
 &=2\{d^*-\mu_1(d^*)\}\{1-\mu_1'(d^*)\}\{\hat d^*-d^*\}-2\{d^*-\mu_1(d^*)\}\{\hat\mu_1(d^*)-\mu_1(d^*)\}\\
 &\quad-\left(\frac{\delta}{\Phi^-(p)}\right)^2\Big\{\{\hat\mu_2(d^*)-\mu_2(d^*)\}-2\mu_1(d^*)\{\hat\mu_1(d^*)-\mu_1(d^*)\}\\
 &\quad+\{\mu_2'(d^*)-2\mu_1(d^*)\mu_1'(d^*)\}\{\hat d^*-d^*\}\Big\}+o_p(1/\sqrt N),
 \end{array}\]
 i.e.,
 \[c_0\{\hat d^*-d^*\}=c_1\{\hat\mu_1(d^*)-\mu_1(d^*)\}+c_2\{\hat\mu_2(d^*)-\mu_2(d^*)\}+o_p(1/\sqrt N),\]
 where
 \[c_0=2\{d^*-\mu_1(d^*)\}\{1-\mu_1'(d^*)\}-\left(\frac{\delta}{\Phi^-(p)}\right)^2\{\mu_2'(d^*)-2\mu_1(d^*)\mu_1'(d^*)\},\]
 \[c_1=2\{d^*-\mu_1(d^*)\}-\left(\frac{\delta}{\Phi^-(p)}\right)^22\mu_1(d^*),~\text{and}~c_2=\left(\frac{\delta}{\Phi^-(p)}\right)^2.\]
 Hence,
 \[\sqrt N\{\hat d^*-d^*\}\overset{d}{\to} N\left(0, \frac 1{c_0^2}(c_1, c_2)\Sigma_0 (c_1, c_2)^{\tau}\right),\]
which implies our main result since $\hat c_0, \hat c_1, \hat c_2, \hat\Sigma_0$ are consistent estimators of $c_0, c_1, c_2, \Sigma_0$, respectively.  The proof is now complete.
 \end{proof}
 
 
\subsection{Standard Deviation Premium Principle}\label{sec_2_3}

We extend the analysis in Section~\ref{sec_2_2} by assuming a decreasing loading factor and the standard deviation principle.  Specifically, a particular choice of $\rho$ is assumed in \eqref{eq:EVP_prob2} as 
\begin{equation}\label{rho_sd}
 \rho=\rho_0 \text{SD}\left(\frac 1N\sum_{i=1}^N(X_i-d)_+\right)=\rho_0N^{-1/2}\sqrt{\nu_2(d)-\nu_1^2(d)}, \end{equation}
 which depends on both $N$ and $d$ and satisfies \eqref{con1}, where
 \[\nu_2(d)=\E\{(X_i-d)_+^2\}=2\int_d^{\infty}\bar F_{X_1}(x)(x-d)\,dx ~\text{satisfying}~\nu_2'(d)=-2\nu_1(d).\]
 Hence, the total cost for the insurer becomes
 \[\tilde T(d)=\sum_{i=1}^N(X_i\wedge d)+N\nu_1(d)+\rho_0\sqrt N \nu_1(d)\sqrt{\nu_2(d)-\nu_1^2(d)},\]
 and the optimal retention should minimize 
 \[\begin{array}{ll}
 \VaR_p(\tilde T(d))&=N\mu_1(d)+\sqrt N\sqrt{\mu_2(d)-\mu_1^2(d)}\Phi^-(p)+o(\sqrt N)\\
 &\quad+N\nu_1(d)+\rho_0\sqrt N \nu_1(d)\sqrt{\nu_2(d)-\nu_1^2(d)}.
 \end{array}\]
Once again, we seek for $d$ that minimizes \eqref{eq:G_SD} below as we ignore the $o(\sqrt N)$ terms:
\begin{equation} \label{eq:G_SD}
\tilde G(d)=N\E(X_1)+\sqrt N\{\Phi^-(p)\sqrt{\mu_2(d)-\mu_1^2(d)}+\rho_0\nu_1(d)\sqrt{\nu_2(d)-\nu_1^2(d)}\}.
\end{equation}
The existence of the approximately optimal retention is shown in
Theorem~\ref{thm:3} below.
\begin{theorem}\label{thm:3} Assume $F_{X_1}(x)$ has the support $[0, \infty)$ (i.e., $F_{X_1}(0)>0$) or $(0, \infty)$, is continuous on $(0, \infty)$, and
 \begin{equation}\label{Th3-1}
 \lim_{t\to\infty}\frac{\bar F_{X_1}(tx)}{\bar F_{X_1}(t)}=x^{-\alpha}~\text{for all}~x>0~\text{and some}~\alpha>2.\end{equation}
If $F_{X_1}(0)>0$, we further assume  that 
\begin{equation}\label{Th3-2}
 \Phi^-(p)\sqrt{\bar F_{X_1}(0)F_{X_1}(0)}<\rho_0\bar F_{X_1}(0)\sqrt{\E(X_1^2)-\{\E(X_1)\}^2}+\rho_0\frac{F_{X_1}(0)\{\E(X_1)\}^2}{\sqrt{\E(X_1^2)-\{\E(X_1)\}^2}}.
\end{equation}
Then, there exists at least one solution of $\tilde G'(d)=0$ and an approximately optimal retention $d^*\in(0, \infty)$ is its smallest solution, which is a local minimum of  $\tilde G(d)$.
\end{theorem}
\begin{proof} Because  \[ \frac{\tilde G'(d)}{\sqrt N}=\Phi^-(p)\frac{\bar F_{X_1}(d) \{d-\mu_1(d)\}}{\sqrt{\mu_2(d)-\mu_1^2(d)}}-\rho_0\bar F_{X_1}(d)\sqrt{\nu_2(d)-\nu_1^2(d)}-\rho_0\frac{F_{X_1}(d)\nu_1^2(d)}{\sqrt{\nu_2(d)-\nu_1^2(d)}}\]
 and
 \begin{equation}\label{pfTh3-1}
 \lim_{d\to0}\frac{\{d-\mu_1(d)\}^2}{\mu_2(d)-\mu_1^2(d)}=\lim_{d\to0}\frac{2\{d-\mu_1(d)\}F_{X_1}(d)}{2\bar F_{X_1}(d)\{d-\mu_1(d)\}}=\frac{F_{X_1}(0)}{\bar F_{X_1}(0)},\end{equation}
 it follows from \eqref{Th3-2} in Theorem~\ref{thm:3} that
 \begin{equation}\label{pfTh3-2}
 \begin{array}{ll}
 \lim_{d\to0}\frac{\tilde G'(d)}{\sqrt N}&=\Phi^-(p)\sqrt{\bar F_{X_1}(0) F_{X_1}(0)}-\rho_0\bar F_{X_1}(0)\sqrt{\E(X_1^2)-\{\E(X_1)\}^2}\\
 &\quad-\rho_0\frac{F_{X_1}(0)\{\E(X_1)\}^2}{\sqrt{\E(X_1^2)-\{\E(X_1)\}^2}}\\
 &<0.\end{array}\end{equation}
 By \eqref{Th3-1}, we have
  \begin{equation}\label{pfTh3-3}\lim_{d\to\infty}\frac{\nu_1(d)}{d\bar F_{X_1}(d)}=\frac 1{\alpha-1}~\text{and}~\lim_{d\to\infty}\frac{\nu_2(d)}{d^2\bar F_{X_1}(d)}=\frac 2{(\alpha-1)(\alpha-2)},\end{equation}
  implying that
  \begin{equation}\label{pfTh3-4}\lim_{d\to\infty}\frac{\tilde G'(d)}{\sqrt N\bar F_{X_1}(d)d}=\frac{\Phi^-(p)}{\sqrt{\E(X_1^2)-\{\E(X_1)\}^2}}>0.\end{equation}
Hence, it follows from \eqref{pfTh3-2} and \eqref{pfTh3-4} that there exists at least one solution of $\tilde G'(d)=0$ for $d\in (0, \infty)$, and let $d^*$ be the smallest solution. Then, there exists $d_1>d^*$ such that $\tilde G'(d)<0$ for $d\in (0, d^*)$, $\tilde G'(d)\ge0$ for $d\in (d^*, d_1)$, and $\tilde G'(d_1)>0$, implying that $d^*$ is a local minimum of $\tilde G(d)$ for $d\in (0, \infty)$.  The proof is now complete.
 \end{proof}

To estimate the optimal retention nonparametrically, we minimize  the following function for $d$:
\begin{equation} \label{eq:G_SD_nonpar}
\hat{\tilde{G}}(d)=\Phi^-(p)\sqrt{\hat{\mu}_2(d)-\hat{\mu}_1^2(d)}+\rho_0\hat{\nu}_1(d)\sqrt{\hat{\nu}_2(d)-\hat{\nu}_1^2(d)},
\end{equation}
where $\hat{\mu}_1(d)$ and $\hat{\mu}_2(d)$ are given by \eqref{eq:mu_hat}, 
\begin{equation} \label{eq:nu_hat}
\hat\nu_1(d)=\frac 1N\sum_{i=1}^N (X_i- d)_{+} ,~\text{and}~\hat\nu_2(d)=\frac 1N\sum_{i=1}^N(X_i-d)_{+}^2.
\end{equation}
Denote $\tilde{d}_{N,\rho_0}$ and $\hat{\tilde{d}}_{N,\rho_0}$ as the minimizers of $\tilde{G}(d)$ in \eqref{eq:G_SD} and $\hat{\tilde{G}}(d)$ in \eqref{eq:G_SD_nonpar}, respectively. Put $\tilde{\bm{Z}}_i(d)=(I(X_i>d),X_i\wedge d,X_i^2\wedge d^2,(X_i-d)_{+},(X_i-d)_{+}^2)^{\tau}$, and let $\tilde{\Sigma}(d)$ be the covariance matrix of $\tilde{\bm{Z}}_i(d)$.  Define
\begin{equation} \label{eq:nu_d_hat}
\hat\nu_1^*(d)=-\frac 1N\sum_{i=1}^NI(X_i>d)~\text{and}~\hat\nu_2^*(d)=-2\hat{\nu}_1(d)
\end{equation}
to estimate $\nu_1'(d)$ and $\nu_2'(d)$ on top of $\hat{\mu}_1^*(d)$ and $\hat{\mu}_2^*(d)$ which are defined in \eqref{eq:mu_d_hat}. We further denote $\hat{\bar{F}}_{X_1}(d)=\sum_{i=1}^NI(X_i>d)/N$ as the empirical survival function of $X_1$ and $\hat{f}_{X_1}(d)$ as any consistent estimator of the density function for $X_1$, e.g., a kernel density estimation.
The asymptotic properties of the approximately optimal retention are provided in Theorem~\ref{thm:4}.
\begin{theorem}\label{thm:4}
Under conditions of Theorem~\ref{thm:3} and \eqref{rho_sd}, and that $X_1$ has a density function $f_{X_1}$, we have
\[\frac{\sqrt N\{\hat{\tilde{d}}^*_{N,\rho_0}-\tilde{d}^*_{N,\rho_0}\}}{\hat b_0^{-1}\sqrt{\hat{\bm{b}}\hat{\tilde{\Sigma}}_0\hat{\bm{b}}^{\tau}}}\overset{d}{\to} N(0, 1),\]
with $\hat{\bm{b}}:=(\hat{b}_1,\hat{b}_2,\hat{b}_3,\hat{b}_4,\hat{b}_5)$, where
\begin{eqnarray*}
&&\hat{b}_0=\frac{\Phi^{-1}(p)\hat{\bar{F}}_{X_1}(\hat{\tilde{d}}^*_{N,\rho_0})}{\sqrt{\hat{\mu}_2(\hat{\tilde{d}}^*_{N,\rho_0})-\hat{\mu}_1^2(\hat{\tilde{d}}^*_{N,\rho_0})}}-\hat{b}_1\hat{f}_{X_1}(\hat{\tilde{d}}^*_{N,\rho_0})+\hat{b}_2\hat{\mu}_1^*(\hat{\tilde{d}}^*_{N,\rho_0})+\hat{b}_3\hat{\mu}_2^*(\hat{\tilde{d}}^*_{N,\rho_0})\\
&&\hspace{2in}+\hat{b}_4\hat{\nu}_1^*(\hat{\tilde{d}}^*_{N,\rho_0})+\hat{b}_5\hat{\nu}_2^*(\hat{\tilde{d}}^*_{N,\rho_0}),\\
&&\hat{b}_1=\Phi^{-1}(p)\frac{\hat{\tilde{d}}^*_{N,\rho_0}-\hat{\mu}_1(\hat{\tilde{d}}^*_{N,\rho_0})}{\sqrt{\hat{\mu}_2(\hat{\tilde{d}}^*_{N,\rho_0})-\hat{\mu}_1^2(\hat{\tilde{d}}^*_{N,\rho_0})}}-\rho_0\frac{\hat{\nu}_2(\hat{\tilde{d}}^*_{N,\rho_0})-2\hat{\nu}_1^2(\hat{\tilde{d}}^*_{N,\rho_0})}{\sqrt{\hat{\nu}_2(\hat{\tilde{d}}^*_{N,\rho_0})-\hat{\nu}_1^2(\hat{\tilde{d}}^*_{N,\rho_0})}},\\
&&\hat{b}_2=\Phi^{-1}(p)\left[-\frac{\hat{\bar{F}}_{X_1}(\hat{\tilde{d}}^*_{N,\rho_0})}{\sqrt{\hat{\mu}_2(\hat{\tilde{d}}^*_{N,\rho_0})-\hat{\mu}_1^2(\hat{\tilde{d}}^*_{N,\rho_0})}}+\frac{\hat{\mu}_1(\hat{\tilde{d}}^*_{N,\rho_0})\hat{\bar{F}}_{X_1}(\hat{\tilde{d}}^*_{N,\rho_0})[\hat{\tilde{d}}^*_{N,\rho_0}-\hat{\mu}_1(\hat{\tilde{d}}^*_{N,\rho_0})]}{(\hat{\mu}_2(\hat{\tilde{d}}^*_{N,\rho_0})-\hat{\mu}_1^2(\hat{\tilde{d}}^*_{N,\rho_0}))^{3/2}}\right],\\
&&\hat{b}_3=-\Phi^{-1}(p)\frac{\hat{\bar{F}}_{X_1}(\hat{\tilde{d}}^*_{N,\rho_0})[\hat{\tilde{d}}^*_{N,\rho_0}-\hat{\mu}_1(\hat{\tilde{d}}^*_{N,\rho_0})]}{2(\hat{\mu}_2(\hat{\tilde{d}}^*_{N,\rho_0})-\hat{\mu}_1^2(\hat{\tilde{d}}^*_{N,\rho_0}))^{3/2}},\\
&&\hat{b}_4=-2\rho_0\hat{\nu}_1(\hat{\tilde{d}}^*_{N,\rho_0})\left[\frac{1-2\hat{\bar{F}}_{X_1}(\hat{\tilde{d}}^*_{N,\rho_0})}{\sqrt{\hat{\nu}_2(\hat{\tilde{d}}^*_{N,\rho_0})-\hat{\nu}_1^2(\hat{\tilde{d}}^*_{N,\rho_0})}}\right.\\
&&\hspace{1.75in}\left.+\frac{\hat{\nu}_2(\hat{\tilde{d}}^*_{N,\rho_0})\hat{\bar{F}}_{X_1}(\hat{\tilde{d}}^*_{N,\rho_0})+\hat{\nu}_1^2(\hat{\tilde{d}}^*_{N,\rho_0})[1-2\hat{\bar{F}}_{X_1}(\hat{\tilde{d}}^*_{N,\rho_0})]}{2(\hat{\nu}_2(\hat{\tilde{d}}^*_{N,\rho_0})-\hat{\nu}_1^2(\hat{\tilde{d}}^*_{N,\rho_0}))^{3/2}}\right],\\
&&\hat{b}_5=\rho_0\left[\frac{\hat{\bar{F}}_{X_1}(\hat{\tilde{d}}^*_{N,\rho_0})}{\sqrt{\hat{\nu}_2(\hat{\tilde{d}}^*_{N,\rho_0})-\hat{\nu}_1^2(\hat{\tilde{d}}^*_{N,\rho_0})}}-\frac{\hat{\nu}_2(\hat{\tilde{d}}^*_{N,\rho_0})\hat{\bar{F}}_{X_1}(\hat{\tilde{d}}^*_{N,\rho_0})\!+\hat{\nu}_1^2(\hat{\tilde{d}}^*_{N,\rho_0})[1\!-\!2\hat{\bar{F}}_{X_1}(\hat{\tilde{d}}^*_{N,\rho_0})]}{2(\hat{\nu}_2(\hat{\tilde{d}}^*_{N,\rho_0})-\hat{\nu}_1^2(\hat{\tilde{d}}^*_{N,\rho_0}))^{3/2}}\right]\!,\\
&&\hat{\tilde{\Sigma}}_0=\frac 1N\sum_{i=1}^N\left[\tilde{\bm{Z}}_i(\hat{\tilde{d}}^*_{N,\rho_0})-\frac{1}{N}\sum_{i'=1}^N\tilde{\bm{Z}}_{i'}(\hat{\tilde{d}}^*_{N,\rho_0})\right]\left[\tilde{\bm{Z}}_i(\hat{\tilde{d}}^*_{N,\rho_0})-\frac{1}{N}\sum_{i'=1}^N\tilde{\bm{Z}}_{i'}(\hat{\tilde{d}}^*_{N,\rho_0})\right]^{\tau}.
\end{eqnarray*}
\end{theorem}
\begin{proof}
For notational convenience, we write $d^*$ and $\hat d^*$ for $\tilde{d}_{N,\rho_0}^*$ and $\hat{\tilde{d}}_{N,\rho_0}^*$, respectively. Then
\begin{equation}\label{normal4}
 \sqrt N
 \begin{pmatrix}
 \hat{\bar{F}}_{X_1}(d^*)-\bar{F}_{X_1}(d^*)\\
 \hat\mu_1(d^*)-\mu_1(d^*)\\
 \hat\mu_2(d^*)-\mu_2(d^*)\\
 \hat\nu_1(d^*)-\nu_1(d^*)\\
 \hat\nu_2(d^*)-\nu_2(d^*)
 \end{pmatrix}\overset{\mathrm{d}}{\to} N(\boldsymbol{0},\tilde{\Sigma}_0)
\end{equation}
when
 $\tilde{\Sigma}(d^*)\to\tilde{\Sigma}_0~\text{as}~N\to\infty$.
Expansion of $\hat{\tilde{G}}'(\hat{d}^*)-\tilde{G}'(d^*)$ yields
\begin{align} \label{eq:thm4_expansion}
0&=\hat{\tilde{G}}'(\hat{d}^*)-\tilde{G}'(d^*)\\
&=\frac{\Phi^{-1}(p)\bar{F}_{X_1}(d^*)}{\sqrt{\mu_2(d^*)-\mu_1^2(d^*)}}\left[\hat{d}^*-d^*\right]+b_1\left[\hat{\bar{F}}_{X_1}(\hat{d}^*)-\bar{F}_{X_1}(d^*)\right]\nonumber\\
&\qquad+b_2\left[\hat{\mu}_1(\hat{d}^*)-\mu_1(d^*)\right]+b_3\left[\hat{\mu}_2(\hat{d}^*)-\mu_2(d^*)\right]\nonumber\\
&\qquad+b_4\left[\hat{\nu}_1(\hat{d}^*)-\nu_1(d^*)\right]+b_5\left[\hat{\nu}_2(\hat{d}^*)-\nu_2(d^*)\right]+o_P(1/\sqrt{N}),\nonumber
\end{align}
where
\begin{eqnarray*}
&& b_1=\Phi^{-1}(p)\frac{{\tilde{d}}^*_{N,\rho_0}-{\mu}_1(d^*)}{\sqrt{{\mu}_2(d^*)-{\mu}_1^2(d^*)}}-\rho_0\frac{{\nu}_2(d^*)-2{\nu}_1^2(d^*)}{\sqrt{{\nu}_2(d^*)-{\nu}_1^2(d^*)}},\\
&& b_2=\Phi^{-1}(p)\left[-\frac{{\bar{F}}_{X_1}(d^*)}{\sqrt{{\mu}_2(d^*)-{\mu}_1^2(d^*)}}+\frac{{\mu}_1(d^*){\bar{F}}_{X_1}(d^*)[{\tilde{d}}^*_{N,\rho_0}-{\mu}_1(d^*)]}{({\mu}_2(d^*)-{\mu}_1^2(d^*))^{3/2}}\right],\\
&& b_3=-\Phi^{-1}(p)\frac{{\bar{F}}_{X_1}(d^*)[{\tilde{d}}^*_{N,\rho_0}-{\mu}_1(d^*)]}{2({\mu}_2(d^*)-{\mu}_1^2(d^*))^{3/2}},\\
&& b_4=-2\rho_0{\nu}_1(d^*)\left[\frac{1-2{\bar{F}}_{X_1}(d^*)}{\sqrt{{\nu}_2(d^*)-{\nu}_1^2(d^*)}}+\frac{{\nu}_2(d^*){\bar{F}}_{X_1}(d^*)+{\nu}_1^2(d^*)[1-2{\bar{F}}_{X_1}(d^*)]}{2({\nu}_2(d^*)-{\nu}_1^2(d^*))^{3/2}}\right],\\
&& b_5=\rho_0\left[\frac{{\bar{F}}_{X_1}(d^*)}{\sqrt{{\nu}_2(d^*)-{\nu}_1^2(d^*)}}-\frac{{\nu}_2(d^*){\bar{F}}_{X_1}(d^*)+{\nu}_1^2(d^*)[1-2{\bar{F}}_{X_1}(d^*)]}{2({\nu}_2(d^*)-{\nu}_1^2(d^*))^{3/2}}\right].
\end{eqnarray*}
We also have
\begin{equation*}\label{eq:thm4_expansion2}
\begin{pmatrix}
 \hat{\bar{F}}_{X_1}(\hat{d}^*)-\bar{F}_{X_1}(d^*)\\
 \hat\mu_1(\hat{d}^*)-\mu_1(d^*)\\
 \hat\mu_2(\hat{d}^*)-\mu_2(d^*)\\
 \hat\nu_1(\hat{d}^*)-\nu_1(d^*)\\
 \hat\nu_2(\hat{d}^*)-\nu_2(d^*)
\end{pmatrix}
=
\begin{pmatrix}
 \hat{\bar{F}}_{X_1}(d^*)-\bar{F}_{X_1}(d^*)\\
 \hat\mu_1(d^*)-\mu_1(d^*)\\
 \hat\mu_2(d^*)-\mu_2(d^*)\\
 \hat\nu_1(d^*)-\nu_1(d^*)\\
 \hat\nu_2(d^*)-\nu_2(d^*)
\end{pmatrix}
+
\begin{pmatrix}
 -f_{X_1}(d^*)\\
 \mu_1'(d^*)\\
 \mu_2'(d^*)\\
 \nu_1'(d^*)\\
 \nu_2'(d^*)
\end{pmatrix}
\left(\hat{d}^*-d^*\right)+o_P(1/\sqrt{N}).
\end{equation*}
The latter together with \eqref{normal4} and \eqref{eq:thm4_expansion} imply that
\begin{equation*}
\sqrt{N}(\hat{d}^*-d^*)=-\sqrt{N}b_0^{-1}\bm{b}
\begin{pmatrix}
 \hat{\bar{F}}_{X_1}(d^*)-\bar{F}_{X_1}(d^*)\\
 \hat\mu_1(d^*)-\mu_1(d^*)\\
 \hat\mu_2(d^*)-\mu_2(d^*)\\
 \hat\nu_1(d^*)-\nu_1(d^*)\\
 \hat\nu_2(d^*)-\nu_2(d^*)
\end{pmatrix}
+o_P(1)\overset{\mathrm{d}}{\to} N(\boldsymbol{0},b_0^{-2}\bm{b}\tilde{\Sigma}_0\bm{b}^{\tau}),
\end{equation*}
where $\bm{b}=(b_1,b_2,b_3,b_4,b_5)$ and 
\begin{eqnarray*}
&&b_0={\Phi^{-1}(p)\bar{F}_{X_1}(d^*)}/{\sqrt{\mu_2(d^*)-\mu_1^2(d^*)}}-b_1f_{X_1}(d^*)+b_2\mu_1'(d^*)+b_3\mu_2'(d^*)\\ &&\hspace{2in} +b_4\nu_1'(d^*)+b_5\nu_2'(d^*).
\end{eqnarray*}
Hence, the theorem follows as $\hat{b}_0$, $\hat{\bm{b}}$ and $\hat{\tilde{\Sigma}}_0$ are consistent estimators of $b_0$, $\bm{b}$ and $\tilde{\Sigma}_0$, respectively.  The proof is now complete.
\end{proof}

\subsection{Sharpe Ratio Premium Principle} \label{sec_2_4}

We now recast the results in Section~\ref{sec_2_3} by assuming the Sharpe Ratio premium principle
 \begin{equation}\label{SR}
 \E\left(\sum_{i=1}^N(X_i-d)_+\right)+\rho_0 \frac{\E(\sum_{i=1}^N(X_i-d)_+)}{\text{SD}(\sum_{i=1}^N(X_i-d)_+)}=N\nu_1(d)+\rho_0\sqrt{N}\frac{\nu_1(d)}{\sqrt{\nu_2(d)-\nu_1^2(d)}}, \end{equation}
  leading to the total cost for the insurer as
 \[\tilde T(d)=\sum_{i=1}^N(X_i\wedge d)+N\nu_1(d)+\rho_0\sqrt N \frac{\nu_1(d)}{\sqrt{\nu_2(d)-\nu_1^2(d)}}.\]
 In this case, the loading factor becomes
 $$\rho=\frac{\rho_0}{\text{SD}(\sum_{i=1}^N(X_i-d)_+)}=\frac{\rho_0}{\sqrt{N}\sqrt{\nu_2(d)-\nu_1^2(d)}},$$
 which is decreasing in $N$ and satisfies \eqref{con1}.
 As before, the optimal retention minimizes 
 \[\begin{array}{ll}
 \VaR_p(\tilde T(d))&=N\mu_1(d)+\sqrt N\sqrt{\mu_2(d)-\mu_1^2(d)}\Phi^-(p)+o(\sqrt N)\\
 &\quad+N\nu_1(d)+\rho_0\sqrt N \frac{\nu_1(d)}{\sqrt{\nu_2(d)-\nu_1^2(d)}},
 \end{array}\]
and we seek for $d$ that minimizes \eqref{eq:G_Sharpe} below by ignoring the $o(\sqrt N)$ terms:
\begin{equation} \label{eq:G_Sharpe}
  \bar G(d)=N\E(X_1)+\sqrt N\left\{\Phi^-(p)\sqrt{\mu_2(d)-\mu_1^2(d)}+\rho_0\frac{\nu_1(d)}{\sqrt{\nu_2(d)-\nu_1^2(d)}}\right\}.
\end{equation}
The existence of the approximately optimal retention is shown in
Theorem~\ref{thm:5} below.
\begin{theorem}\label{thm:5} Assume $F_{X_1}(x)$ has the support $[0, \infty)$ (i.e., $F_{X_1}(0)>0$) or $(0, \infty)$, is continuous on $(0, \infty)$, and 
\begin{equation}\label{Th5-1}
 \lim_{t\to\infty}\frac{\bar F_{X_1}(tx)}{\bar F_{X_1}(t)}=x^{-\alpha}~\text{for all}~x>0~\text{and some}~\alpha\in (2, 4).\end{equation}
If $F_{X_1}(0)>0$, we further assume  that 
 \begin{equation}\label{Th5-2}
 \Phi^-(p)\sqrt{\bar F_{X_1}(0)F_{X_1}(0)}<\rho_0\frac{\bar F_{X_1}(0)}{\sqrt{\E(X_1^2)-\{\E(X_1)\}^2}}-\rho_0\frac{F_{X_1}(0)\{\E(X_1)\}^2}{\{\E(X_1^2)-\{\E(X_1)\}^2\}^{3/2}}.\end{equation}
Then, there exists at least one solution of $\bar G'(d)=0$ and an approximately optimal retention $d^*\in(0, \infty)$ is its smallest solution, which is a local minimum of  $\bar G(d)$.
\end{theorem}
\begin{proof} Because  \[ \frac{\bar G'(d)}{\sqrt N}=\Phi^-(p)\frac{\bar F_{X_1}(d) \{d-\mu_1(d)\}}{\sqrt{\mu_2(d)-\mu_1^2(d)}}-\rho_0\frac{\bar F_{X_1}(d)}{\sqrt{\nu_2(d)-\nu_1^2(d)}}+\rho_0\frac{F_{X_1}(d)\nu_1^2(d)}{\{\nu_2(d)-\nu_1^2(d)\}^{3/2}},\]
 it follows from \eqref{Th3-1} and \eqref{Th5-2} that
 \begin{equation}\label{pfTh5-1}
 \begin{array}{ll}
 \lim_{d\to0}\frac{\tilde G'(d)}{\sqrt N}&=\Phi^-(p)\sqrt{\bar F_{X_1}(0) F_{X_1}(0)}-\rho_0\frac{\bar F_{X_1}(0)}{\sqrt{\E(X_1^2)-\{\E(X_1)\}^2}}\\
 &\quad+\rho_0\frac{F_{X_1}(0)\{\E(X_1)\}^2}{\{\E(X_1^2)-\{\E(X_1)\}^2\}^{3/2}}\\
 &<0.\end{array}\end{equation}
 By \eqref{Th5-1} with $\alpha<4$ and \eqref{pfTh3-3}, 
   \begin{equation}\label{pfTh5-2}
   \lim_{d\to\infty}\frac{\bar G'(d)}{\sqrt N\bar F_{X_1}(d)d}=\frac{\Phi^-(p)}{\sqrt{\E(X_1^2)-\{\E(X_1)\}^2}}>0.\end{equation}
 Hence, it follows from \eqref{pfTh5-1} and \eqref{pfTh5-2} that there exists at least one solution of $\bar G'(d)=0$ for $d\in (0, \infty)$, and let $d^*$ be the smallest solution. Then, there exists $d_1>d^*$ such that $\bar G'(d)<0$ for $d\in (0, d^*)$, $\bar G'(d)\ge0$ for $d\in (d^*, d_1)$, and $\bar G'(d_1)>0$, implying that $d^*$ is a local minimum of $\bar G(d)$ for $d\in (0, \infty)$.  The proof is now complete.
 \end{proof}

To estimate the optimal retention nonparametrically, we minimize the following function for $d$:
\begin{equation} \label{eq:G_Sharpe_nonpar}
\hat{\bar{G}}(d)=\Phi^-(p)\sqrt{\hat{\mu}_2(d)-\hat{\mu}_1^2(d)}+\rho_0\frac{\hat{\nu}_1(d)}{\sqrt{\hat{\nu}_2(d)-\hat{\nu}_1^2(d)}}.
\end{equation}
Denote $\bar{d}_{N,\rho_0}$ and $\hat{\bar{d}}_{N,\rho_0}$ as the minimizers of $\bar{G}(d)$ in \eqref{eq:G_Sharpe} and $\hat{\bar{G}}(d)$ in \eqref{eq:G_Sharpe_nonpar}, respectively. The asymptotic properties of the approximately optimal retention are provided in Theorem~\ref{thm:6}.
\begin{theorem}\label{thm:6}
Under conditions of Theorem~\ref{thm:5} and \eqref{SR}, and that $X_1$ has a density function $f_{X_1}$, we have
\[\frac{\sqrt N\{\hat{\bar{d}}^*_{N,\rho_0}-\bar{d}^*_{N,\rho_0}\}}{\hat a_0^{-1}\sqrt{\hat{\bm{a}}\hat{\bar{\Sigma}}_0\hat{\bm{a}}^{\tau}}}\overset{d}{\to} N(0, 1),\]
with $\hat{\bm{a}}:=(\hat{a}_1,\hat{a}_2,\hat{a}_3,\hat{a}_4,\hat{a}_5)$, where $\hat{a}_2$, $\hat{a}_3$ and $\hat{\bar{\Sigma}}_0$ are identical to $\hat{b}_2$, $\hat{b}_3$ and $\hat{\tilde{\Sigma}}_0$, respectively, in Theorem \ref{thm:4} though $\hat{\tilde{d}}^*_{N,\rho_0}$ is replaced by $\hat{\bar{d}}^*_{N,\rho_0}$, and
\begin{eqnarray*}
&& \hat{a}_0=\frac{\Phi^{-1}(p)\hat{\bar{F}}_{X_1}(\hat{\bar{d}}^*_{N,\rho_0})}{\sqrt{\hat{\mu}_2(\hat{\bar{d}}^*_{N,\rho_0})-\hat{\mu}_1^2(\hat{\bar{d}}^*_{N,\rho_0})}}-\hat{a}_1\hat{f}_{X_1}(\hat{\bar{d}}^*_{N,\rho_0})+\hat{a}_2\hat{\mu}_1^*(\hat{\bar{d}}^*_{N,\rho_0})+\hat{a}_3\hat{\mu}_2^*(\hat{\bar{d}}^*_{N,\rho_0})\\
&&\hspace{2in}+\hat{a}_4\hat{\nu}_1^*(\hat{\bar{d}}^*_{N,\rho_0})+\hat{a}_5\hat{\nu}_2^*(\hat{\bar{d}}^*_{N,\rho_0}),\\
&&\hat{a}_1=\Phi^{-1}(p)\frac{\hat{\tilde{d}}^*_{N,\rho_0}-\hat{\mu}_1(\hat{\bar{d}}^*_{N,\rho_0})}{\sqrt{\hat{\mu}_2(\hat{\bar{d}}^*_{N,\rho_0})-\hat{\mu}_1^2(\hat{\bar{d}}^*_{N,\rho_0})}}-\rho_0\frac{\hat{\nu}_2(\hat{\bar{d}}^*_{N,\rho_0})}{(\hat{\nu}_2(\hat{\bar{d}}^*_{N,\rho_0})-\hat{\nu}_1^2(\hat{\bar{d}}^*_{N,\rho_0}))^{3/2}},\\
&&\hat{a}_4=\rho_0\hat{\nu}_1(\hat{\bar{d}}^*_{N,\rho_0})\left[\frac{2}{(\hat{\nu}_2(\hat{\bar{d}}^*_{N,\rho_0})-\hat{\nu}_1^2(\hat{\bar{d}}^*_{N,\rho_0}))^{3/2}}+\frac{3[\hat{\nu}_1^2(\hat{\bar{d}}^*_{N,\rho_0})-\hat{\nu}_2(\hat{\bar{d}}^*_{N,\rho_0})\hat{\bar{F}}_{X_1}(\hat{\bar{d}}^*_{N,\rho_0})]}{(\hat{\nu}_2(\hat{\bar{d}}^*_{N,\rho_0})-\hat{\nu}_1^2(\hat{\bar{d}}^*_{N,\rho_0}))^{5/2}}\right],\\
&&\hat{a}_5=-\rho_0\left[\frac{\hat{\bar{F}}_{X_1}(\hat{\bar{d}}^*_{N,\rho_0})}{(\hat{\nu}_2(\hat{\bar{d}}^*_{N,\rho_0})-\hat{\nu}_1^2(\hat{\bar{d}}^*_{N,\rho_0}))^{3/2}}+\frac{3[\hat{\nu}_1^2(\hat{\bar{d}}^*_{N,\rho_0})-\hat{\nu}_2(\hat{\bar{d}}^*_{N,\rho_0})\hat{\bar{F}}_{X_1}(\hat{\bar{d}}^*_{N,\rho_0})]}{2(\hat{\nu}_2(\hat{\bar{d}}^*_{N,\rho_0})-\hat{\nu}_1^2(\hat{\bar{d}}^*_{N,\rho_0}))^{5/2}}\right].
\end{eqnarray*}
\end{theorem}
\begin{proof}
Since the proof is similar to that for Theorem \ref{thm:4}, we omit the details.
\end{proof}

  \begin{remark}\label{rem1}
Another choice of the reinsurance premium beyond the Sharpe Ratio may also seem natural. That is, we could use the Standard Deviation to determine the reinsurance premium as follows:
 \[\E\left(\sum_{i=1}^N(X_i-d)_+\right)+\rho_0\text{SD}\left(\sum_{i=1}^N(X_i-d)_+\right)=N\nu_1(d)+\rho_0\sqrt{N}\sqrt{\nu_2(d)-\nu_1^2(d)}.\]
However, the resulting objective function has a positive derivative at $d=0$, which often leads to a trivial approximately optimal retention being either zero or infinity. Therefore, we do not discuss this setting in the paper as the optimal retention is trivial.
\end{remark}

\section{Generalization to distortion risk measures}
\label{sec:es}


We show now that our results in Section~\ref{sec:var} can be naturally extended to optimal reinsurance problems under general distortion risk measures. A large class of quantile-based risk measures is the distorted class, for which Definition~\ref{dist_def} is needed. 
\begin{definition}\label{dist_def}
A distortion function is a non-decreasing function $h: [0,1]\to[0,1]$ such that $h(0)=h(0+)=0$ and $h(1)=h(1-)=1$.
\end{definition}
Yarri's dual theory of choice under risk -- e.g., see Yaari (1987) -- postulates that the risk preferences of a nonrisk neutral decision maker could be modeled by an expectation concerning a reweighed or distorted probability measure, where the distortion function is as in Definition~\ref{dist_def}. We are ready to define a \emph{distortion risk measure}, which is given as Definition~\ref{dist_RM_def}. 
\begin{definition}\label{dist_RM_def}
Let $Y$ be a non-negative random variable and $h$ be 
a distortion function. The Choquet integral
\begin{eqnarray}
\varphi_h(Y):=\int_0^\infty h\circ \bar{F}_Y(x)\;dx=\int_0^\infty \big(1-\tilde{h}\circ F_Y(x)\big)\;dx
\end{eqnarray}
is called a distortion risk measure, where $\tilde{h}(\cdot)=1-h(1-\cdot)$ on $[0,1]$.
\end{definition}
Note that $\tilde{h}$ is a distortion function since $h$ is a distortion function. Further, $\rho_h(Y)$ is an expectation with respect to a reweighed probability measure, namely $\tilde{h}\circ F_Y$; that is, $\rho_h(Y)=\int_0^\infty x \;d\tilde{h}\circ F_Y(x)$ is a Lebesgue-Stieljes integral.  It is not difficult to see that VaR and ES are distortion risk measures with distortion functions $h_{\VaR_p}(s):=I_{\{p\le s\le1\}}$ and $h_{\ES_p}(s):=\min\left(\frac{s}{1-p},1\right)$, respectively for all $s\in[0,1]$. Other examples are i) \emph{Dual-power} with $h_{DP}(s):=1-(1-s)^\beta, \beta\ge1$, ii)
\emph{Gini} with $h_{G}(s):=(1+\beta)s-\beta s^2, 0\le\beta\le1$,
iii) \emph{Proportional hazard transform (PHT)} with $h_{PHT}(s):= s^{1-\beta}, 0\le \beta<1$, and iv) \emph{Wang transform} with $h_{WT}(s):= \Phi\big(\Phi^-(s)+\beta\big), \beta\ge 0$.

The results in Section~\ref{sec:var} could be generalized to the class of distortion risk measures through Lemma~\ref{lem:robust} after verifying some robustness conditions.  Note that the case in which the risk preferences are ordered by ES is a special case of our main results in this section.

\begin{lemma}\label{lem:robust}
Let $Z$ follow the standard normal distribution $\mathrm{N}(0,1)$. We have
\[\varphi_{h}\big(T(d,N,\rho)\big)=N(\E(X_1)+\rho\nu_1(d))+\sqrt{N}\sqrt{\mu_2(d)-\mu^2_1(d)}\varphi_h(Z)+o(\sqrt{N}).\]
\end{lemma}
\begin{proof}
It follows from the \emph{Central Limit Theorem} that     
$$F^-_{T(d,N,\rho)}(p)=N(\E(X_1)+\rho\nu_1(d))+\sqrt{N}\sqrt{\mu_2(d)-\mu^2_1(d)}\Phi^-(\alpha)+o(\sqrt{N}),~~p\in(0,1).$$As $0\le \sum^n_{i=1}(X_i\wedge d)\le Nd$ and $Z$ is integrable, we have $\{Z\}\cup\{T(d,N,\rho):N=1,2\dots\}$ is $h$-uniformly integrable.\footnote{For a distortion function $h$, a set of random variables $\mathcal X$ is called \emph{$h$-uniformly integrable} if
$$\lim_{k\downarrow 0}\sup_{S\in\mathcal{X}}\int^k_0|F^{-}_S(1-t)|\, d h(t)=0~~\text{and}~~\lim_{k\uparrow 1}\sup_{S\in\mathcal{X}}\int^1_k|F^{-}_S(1-t)|\, dh(t)=0.$$}  By Theorem~4 of Wang et al. (2020), translation invariance and homogeneity of $\varphi_h$ of order 1,
\[\varphi_{h}\big(T(d,N,\rho)\big)=N\big(\E(X_1)+\rho\nu_1(d)\big)+\sqrt{N}\sqrt{\mu_2(d)-\mu^2_1(d)}\varphi_h(Z)+o(\sqrt{N}).\qedhere\]
   \end{proof}

Therefore, instead of minimizing $\varphi_{h}\big(T(d,N,\rho)\big)$ to define the optimal retention $d$, we seek an approximately optimal retention $d$ by minimizing 
\begin{equation}\label{eq:EVP_prob1}
    G_{\varphi}(d):=N\E(X_1)+N\rho\nu_1(d)+\sqrt{N}\sqrt{\mu_2(d)-\mu^2_1(d)}\varphi_h(Z).
\end{equation}
Similar to Theorem~\ref{thm:1}, we can show the unique solution after replacing $\Phi^-(p)$ in Theorem~\ref{thm:1} with $\varphi_h(Z)$. Further, we can estimate this unique approximately optimal retention and derive its asymptotic normal limit.
Specifically, we have a generalized result below following a proof similar to that of Theorem~\ref{thm:1}, which is given as Theorem~\ref{thm:general} that needs the following notation. 
$$H_{\varphi}(d):=\{d-\mu_1(d)\}^2-\left(\frac{\sqrt N\rho}{\varphi_h(Z)}\right)^2\{\mu_2(d)-\mu_1^2(d)\}=0.$$

\begin{theorem}\label{thm:general}
Assume $\E(X_1)<\infty$, $F_{X_1}(x)$ has the support $[0, \infty)$ (i.e., $F_{X_1}(0)>0$) or $(0, \infty)$, and is continuous on $(0, \infty)$. When the support is $[0, \infty)$, we further assume 
$F_{X_1}(0)<\frac{N\rho^2}{N\rho^2+(\varphi_h(Z))^2}$, which is always true when $N$ is large enough. Then, there exists a unique approximately optimal retention $d^*_{\varphi,N}\in(0, \infty)$ such that 
\[d^*_{\varphi,N}=\argmin_{d>0}G_{\varphi}(d)~~\text{and}~~H_{\varphi}(d_{\varphi,N}^*)=0.\]
{}
\end{theorem}
In light of Lemma~\ref{lem:robust}, we can also extend Theorems~\ref{thm:2}-\ref{thm:4} in a similar sense to Theorem~\ref{thm:1} by changing $\Phi^-(p)$ to $\varphi_h(Z)$ for which no other adjustments are needed.

\section{Simulation studies}\label{sec_sim}

We conduct two simulation studies in this section.
Section~\ref{sec:sim1} assesses the validity of substituting the actual VaR of total cost $\VaR_p\big(T(d,N,\rho)\big)$ as specified in \eqref{T_2_EVP} with the normal-approximated VaR $G_{N,\rho}(d)$ outlined in \eqref{eq:EVP_prob2}; this assessment is conducted under the four loading factor rules delineated in Sections \ref{sec_2_1}--\ref{sec_2_4}. Section~\ref{sec:sim2} empirically examines the statistical properties of the optimal retention estimators introduced in Theorems~\ref{thm:2}, \ref{thm:4}, and \ref{thm:6}.

\subsection{Examining validity of approximately optimal retention} \label{sec:sim1}

We generate samples of $X_i$ from a Pareto (type-II) distribution with a probability density function given by $f_X(x) = (\alpha/\lambda)(1+x/\lambda)^{-(\alpha+1)}$, where the shape parameter $\alpha=9$ and the scale parameter $\lambda=8$, such that $\mathbb{E}[X_i] = \alpha/(\lambda-1) = 1$. We set the risk level at $p=0.75$ and consider sample sizes of $N=10$, $25$, and $100$. In this analysis, we examine all four loading factor rules by setting $\rho=0.3$ for the constant loading factor,  choosing $\delta=0.5$ for the decreasing loading factor, and $\rho_0=0.5$ for the remaining two rules.

We compute the true VaR of the total cost $\text{VaR}_p\big(T(d,N,\rho)\big)$ using \eqref{T_2_EVP}, where $\mathbb{E}\{(X_1-d)_{+}\}$ and $\text{VaR}_p\allowbreak(\sum_{i=1}^{N}(X_i\wedge d))$ are approximated 
through $B=50000$ simulated samples. For example, we have $\mathbb{E}\{\sum_{i=1}^{N}(X_i\wedge d)\}\approx (1/B)\sum_{j=1}^{B}\sum_{i=1}^{N}(X_{ij}\wedge d)$ with $X_{ij}$ iid sampled from a Pareto distribution for $i=1,\ldots,N$ and $j=1,\ldots,B$. We also compute the normal-approximated VaR of total cost $G_{N,\rho}(d)$ in \eqref{eq:EVP_prob2} by numerical integration.  

\begin{figure}[!ht]
\begin{center}
\begin{subfigure}[h]{0.49\linewidth}
\includegraphics[width=\linewidth]{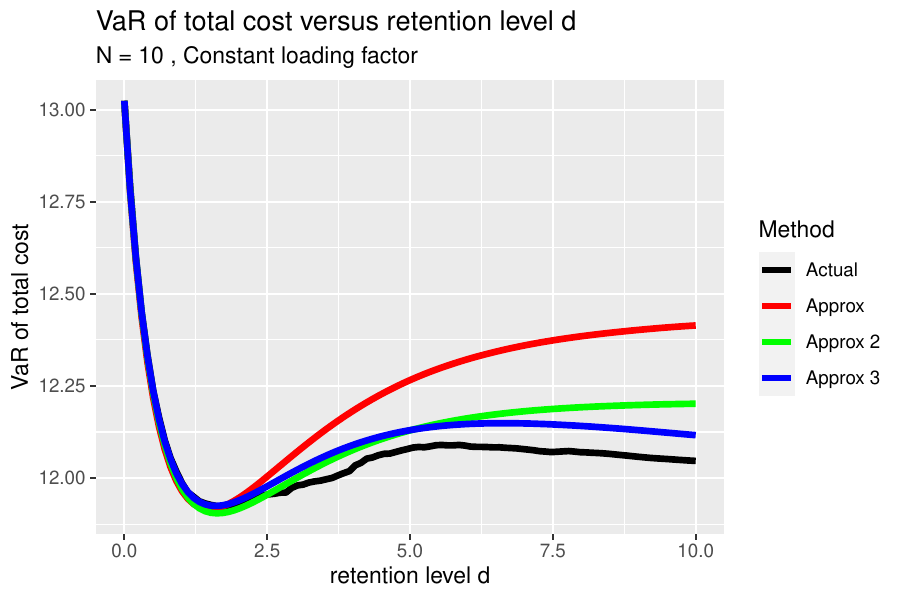}
\end{subfigure}
\begin{subfigure}[h]{0.49\linewidth}
\includegraphics[width=\linewidth]{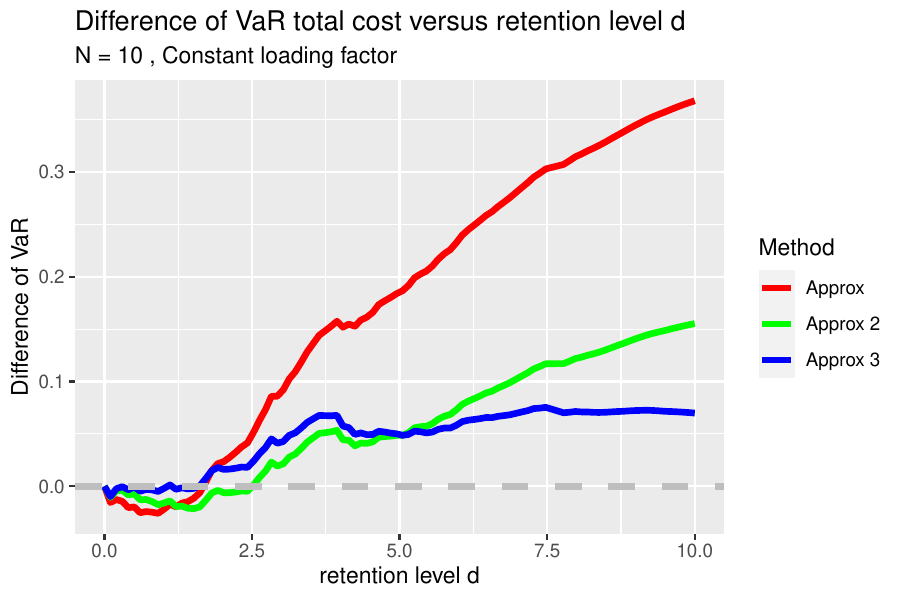}
\end{subfigure}
\begin{subfigure}[h]{0.49\linewidth}
\includegraphics[width=\linewidth]{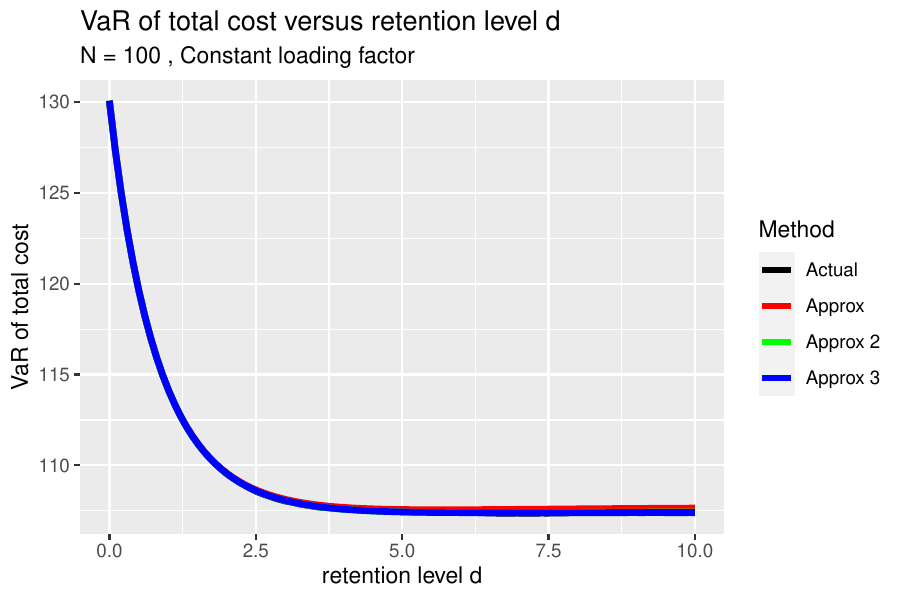}
\end{subfigure}
\begin{subfigure}[h]{0.49\linewidth}
\includegraphics[width=\linewidth]{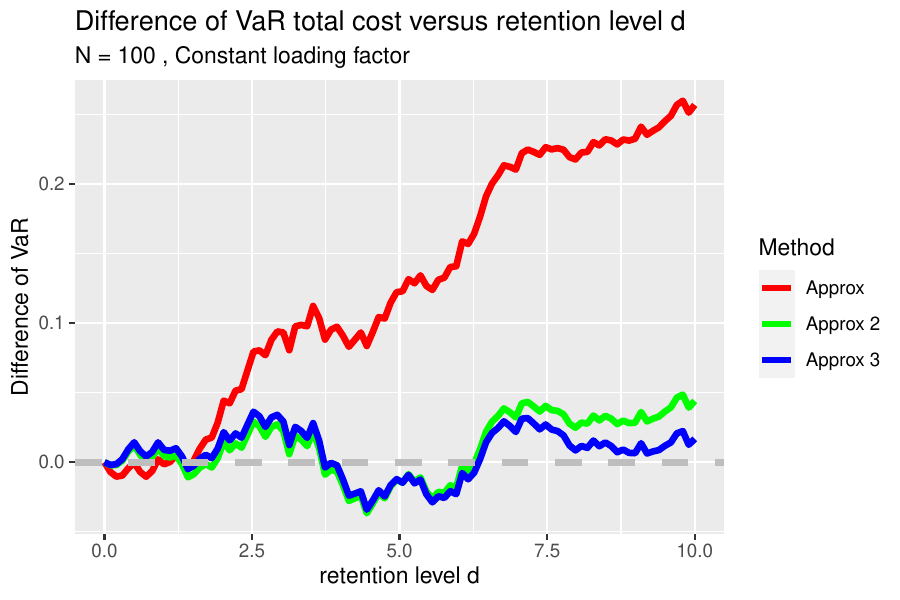}
\end{subfigure}
\end{center}
\vspace{-1.5em}
\caption{\emph{Left panels}: $\text{VaR}_p\big(T(d,N,\rho)\big)$ (black curves), $G_{N,\rho}(d)$ (red curves), $G_{N,\rho}^{(2)}(d)$ (green curves) and $G_{N,\rho}^{(3)}(d)$ (blue curves) for $N=10,100$ under the constant loading factor. \emph{Right panels}: $[G_{N,\rho}(d)-\text{VaR}_p\big(T(d,N,\rho)\big)]$ (red curves), $[G_{N,\rho}^{(2)}(d)-\text{VaR}_p\big(T(d,N,\rho)\big)]$ (green curves) and $[G_{N,\rho}^{(3)}(d)-\text{VaR}_p\big(T(d,N,\rho)\big)]$ (blue curves) versus $d$.}
\label{fig:sim1_0}
\end{figure}

\begin{figure}[!h]
\begin{center}
\begin{subfigure}[h]{0.49\linewidth}
\includegraphics[width=\linewidth]{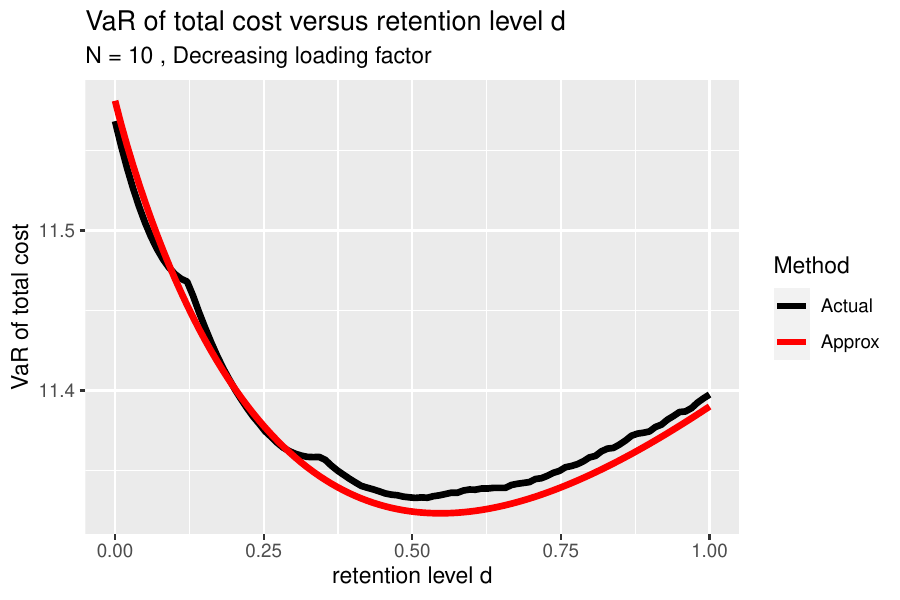}
\end{subfigure}
\begin{subfigure}[h]{0.49\linewidth}
\includegraphics[width=\linewidth]{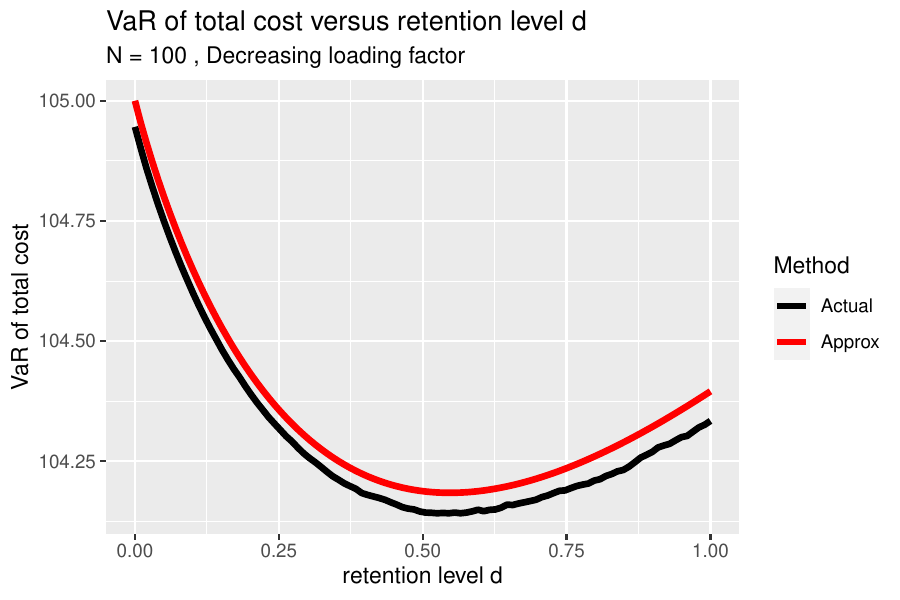}
\end{subfigure}
\begin{subfigure}[h]{0.49\linewidth}
\includegraphics[width=\linewidth]{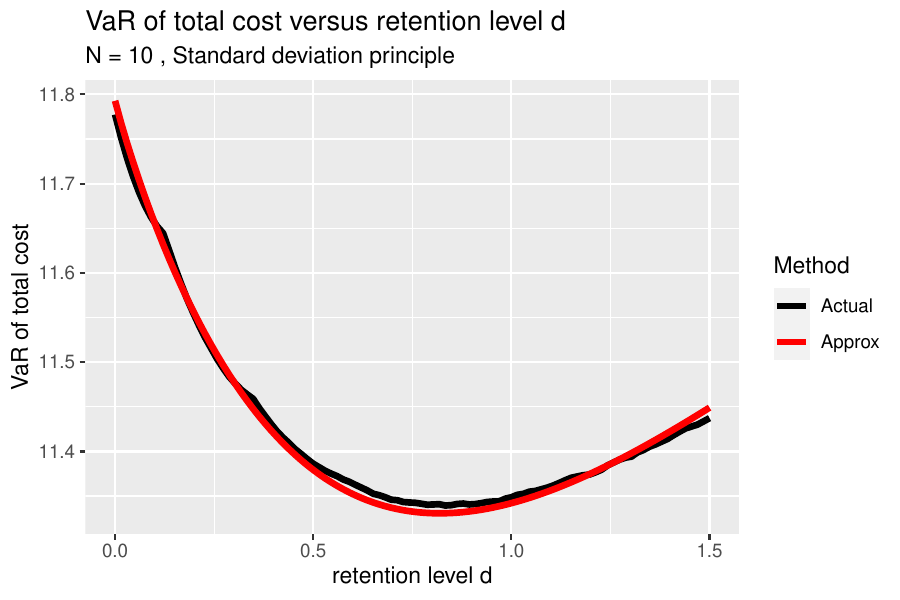}
\end{subfigure}
\begin{subfigure}[h]{0.49\linewidth}
\includegraphics[width=\linewidth]{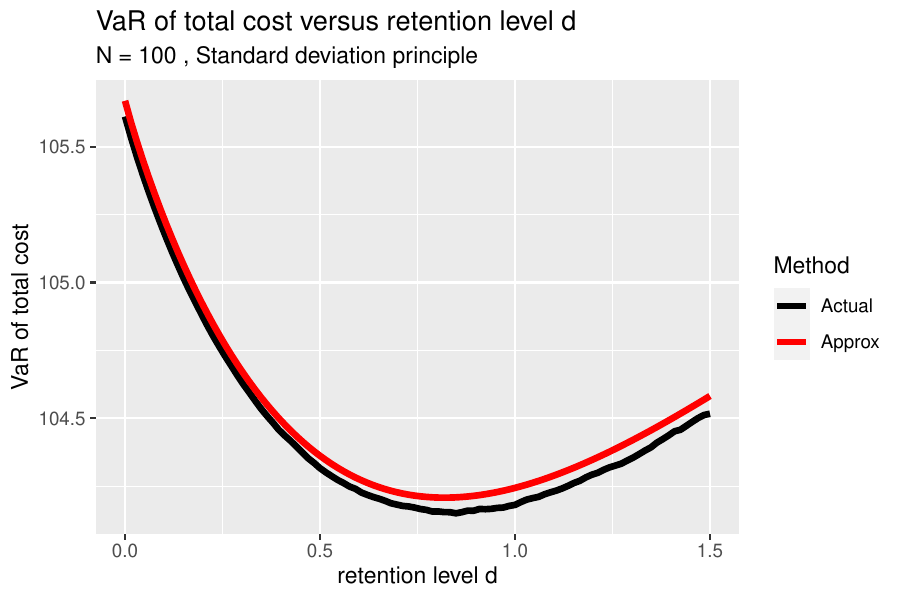}
\end{subfigure}
\begin{subfigure}[h]{0.49\linewidth}
\includegraphics[width=\linewidth]{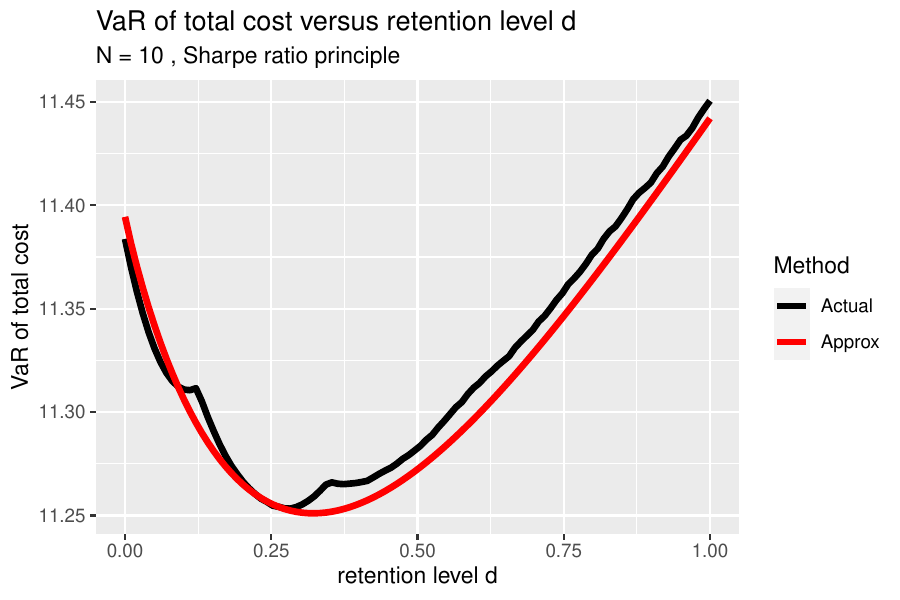}
\end{subfigure}
\begin{subfigure}[h]{0.49\linewidth}
\includegraphics[width=\linewidth]{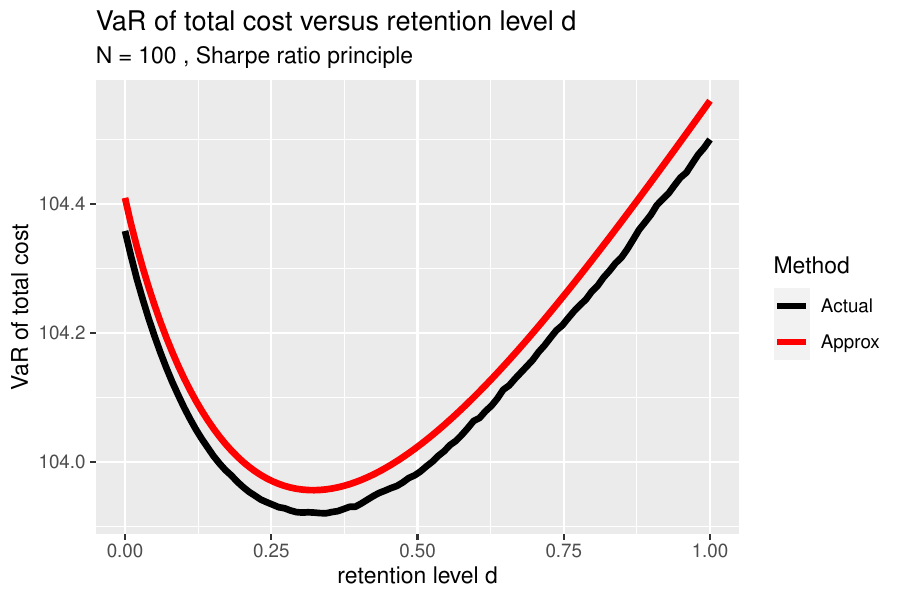}
\end{subfigure}
\end{center}
\vspace{-1.5em}
\caption{$\text{VaR}_p\big(T(d,N,\rho)\big)$ (black curves) and $G_{N,\rho}(d)$ (red curves) versus or $N=10,100$ under various loading factor rules.}
\label{fig:sim1_1}
\end{figure}

Figure~\ref{fig:sim1_0} (left panel) and Figure~\ref{fig:sim1_1} plot $G_{N,\rho}(d)$ (red curves) and $\text{VaR}_p\big(T(d,N,\rho)\big)$ (black curves) as a function of $d$ for $N=10$ and $100$ under the four loading factor rules. We note that $G_{N,\rho}(d)$ approximates closely to $\text{VaR}_p\big(T(d,N,\rho)\big)$ under all loading factor rules. For all loading factor rules except for the constant loading factor, we observe an optimal retention $d^*\in[0,1]$ that minimizes the VaR of the total cost. Conversely, for the constant loading factor, both $\text{VaR}_p\big(T(d,N,\rho)\big)$ and $G_{N,\rho}(d)$ become flat as $d$ increases, especially when $N$ is large. Hence, it is difficult to identify $d^*$ by solely observing the plots.

We then numerically calculate the actual and approximately optimal retentions $d^*$ and $d^*_{N,\rho}$, and the results are displayed in Table \ref{tab:sim1}. Recall that the approximation order of $G_{N,\rho}(d)$ in \eqref{eq:EVP_prob2} is $o(\sqrt{N})$. For all loading factor rules other than the constant loading factor, the actual optimal retention $d^*$, which minimizes the true VaR of total cost $\text{VaR}_p\big(T(d,N,\rho)\big)$, yields similar values as the approximately optimal retention $d^*_{N,\rho}$, which minimizes the approximated VaR of total cost $G_{N,\rho}(d)$. Also, the relative difference $|(d^*_{N,\rho}-d^*)/d^*_{N,\rho}|$ generally reduces as $N$ increases. Furthermore, $d^*$ does not change substantially as $N$ changes. As a result, the approximate optimal retention approach for VaR works well under these three loading factor rules.

\begin{table}[!h]
\small
\caption{\label{tab:sim1} Actual optimal retention $d^*$, approximately optimal retention $d^*_{N,\rho}$, and the relative difference between $d^*$ and $d^*_{N,\rho}$ (in \%) across various loading factor rules, $N$, and approximation orders.}
\centering
\begin{tabular}{crcrrr}
\toprule
Loading factor rule & \multicolumn{1}{c}{$N$} & Approx. order & \multicolumn{1}{c}{Actual} & \multicolumn{1}{c}{Approx.} & \multicolumn{1}{c}{Diff. (\%)} \\ \hline
Constant loading factor & 10 & $o(\sqrt{N})$ & 1.8549 & 1.4856 & -19.91 \\
Constant loading factor & 10 & $o(1)$ & 1.8549 & 1.6276 & -12.25 \\
Constant loading factor & 10 & $o(1/\sqrt{N})$ & 1.8549 & 1.5921 & -14.17 \\ \hline
Constant loading factor & 25 & $o(\sqrt{N})$ & 3.4442 & 2.6838 & -22.08 \\
Constant loading factor & 25 & $o(1)$ & 3.4442 & 2.9634 & -13.96 \\
Constant loading factor & 25 & $o(1/\sqrt{N})$ & 3.4442 & 2.9969 & -12.99 \\ \hline
Constant loading factor & 100 & $o(\sqrt{N})$ & 7.1241 & 5.6581 & -20.58 \\
Constant loading factor & 100 & $o(1)$ & 7.1241 & 6.3361 & -11.06 \\
Constant loading factor & 100 & $o(1/\sqrt{N})$ & 7.1241 & 6.6660 & -6.43 \\ \hline
Decreasing loading factor & 10 & $o(\sqrt{N})$ & 0.5034 & 0.5472 & 8.70 \\
Decreasing loading factor & 25 & $o(\sqrt{N})$ & 0.5835 & 0.5472 & -6.21 \\
Decreasing loading factor & 100 & $o(\sqrt{N})$ & 0.5472 & 0.5472 & 0.02 \\ \hline
Standard deviation principle & 10 & $o(\sqrt{N})$ & 0.7847 & 0.8189 & 4.36 \\
Standard deviation principle & 25 & $o(\sqrt{N})$ & 0.8187 & 0.8189 & 0.03 \\
Standard deviation principle & 100 & $o(\sqrt{N})$ & 0.8499 & 0.8189 & -3.64 \\ \hline
Sharpe ratio principle & 10 & $o(\sqrt{N})$ & 0.2797 & 0.3218 & 15.06 \\
Sharpe ratio principle & 25 & $o(\sqrt{N})$ & 0.3149 & 0.3218 & 2.18 \\
Sharpe ratio principle & 100 & $o(\sqrt{N})$ & 0.3203 & 0.3218 & 0.45\\
\hhline{======}
\end{tabular}
\end{table}

For the constant loading factor, while the optimal retention exists despite the flatness of the curves in the left panels of Figure \ref{fig:sim1_0}, there are noticeable discrepancies between $d^*$ and $d^*_{N,\rho}$ for a given $N$, especially when $N$ is large. Hence, the normal approximation with order $o(\sqrt{N})$ is not sufficiently accurate in determining the optimal retention under this loading factor rule. One may address this issue by employing Edgeworth expansion to improve the approximation precision. Denote $G_{N,\rho}^{(2)}$ and $G_{N,\rho}^{(3)}$ as the Edgeworth approximated VaRs with orders $o(1)$ and $o(1/\sqrt{N})$, respectively; see Section~\ref{apx:sec:edgeworth} of the supplementary material for more details. We now add $G_{N,\rho}^{(2)}$ (green curves) and $G_{N,\rho}^{(3)}$ (blue curves) to the left panels of Figure \ref{fig:sim1_0} and the approximation errors $[G_{N,\rho}^{(2)}-\text{VaR}_p\big(T(d,N,\rho)\big)]$ (green curves) and $[G_{N,\rho}^{(3)}-\text{VaR}_p\big(T(d,N,\rho)\big)]$ (blue curves) to the right panels. We numerically calculate the optimal retention that minimizes the approximated VaRs $G_{N,\rho}^{(2)}(d)$ and $G_{N,\rho}^{(3)}(d)$, and the results are added to Table \ref{tab:sim1}. From the right panel, it is apparent that the approximation errors of VaR, especially for the higher-order Edgeworth expansions (blue curves), are significantly reduced compared to the normal approximations. The discrepancies between $d^*$ and $d^*_{N,\rho}$ are also significantly reduced by employing higher-order approximation methods. In a nutshell, the use of a constant loading factor needs a higher order Edgeworth expansion for the approximately optimal retention when $N$ is large.

\subsection{Verifying statistical properties of nonparametric approach} \label{sec:sim2}
We adopt the same simulation setup as in Section~\ref{sec:sim1} except assessing larger sample sizes of $N=500,$ $2000,$ and $10000$. The computational complexity associated with such large sample sizes is given by the fact the direct estimation of the true VaR of the total cost $\text{VaR}_p\big(T(d,N,\rho)\big)$ is unfeasible without resorting to the normal approximation technique. In each simulation run, we compute the nonparametric estimation of the approximately optimal retention, which is $\hat{d}^*_{N,\rho_N}$ under the decreasing loading factor, $\hat{\tilde{d}}_{N,\rho_N}$ under the standard deviation principle, or $\hat{\bar{d}}_{N,\rho_N}$ under the Sharpe ratio principle. To obtain an $M$-vector of estimated optimal retentions, which is $\{\hat{d}^{*(m)}_{N,\rho_N}\}_{m=1,\ldots,M}$ for each of the three loading factor rules, we repeat the simulation runs $M=5000$ times. Additionally, through numerical integration, we compute the ``true'' approximately optimal retention, denoted as $d^*_{N,\rho_N}$ under the decreasing loading factor, $\tilde{d}_{N,\rho_N}$ under the standard deviation principle, or $\bar{d}_{N,\rho_N}$ under the Sharpe ratio principle. We calculate the sample mean of $\{\hat{d}^{*(m)}_{N,\rho_N}\}_{m=1,\ldots,M}$ and compare it with the true approximately optimal retention to evaluate the bias of the proposed nonparametric estimation approach. 

\begin{table}[!h]
\small
\caption{\label{tab:sim2} \emph{Columns 2--4}: True approximately optimal retention, the sample mean of the nonparametrically estimated approximately optimal retentions, and their relative difference; \emph{Columns 5--7}: Theoretical and empirical standard error of the estimated optimal retention, and their relative difference.}
\centering
\begin{tabular}{lrrrrrr}
\toprule
 & \multicolumn{3}{c}{Mean optimal retention} & \multicolumn{3}{c}{Std. Error optimal retention} \\
\cmidrule(l{3pt}r{3pt}){2-4} \cmidrule(l{3pt}r{3pt}){5-7}
 & \multicolumn{1}{c}{True} & \multicolumn{1}{c}{Estimated} & \multicolumn{1}{c}{Bias (\%)} & \multicolumn{1}{c}{Theoretical} & \multicolumn{1}{c}{Empirical} & \multicolumn{1}{c}{Diff. (\%)} \\ \hline
\multicolumn{7}{l}{Decreasing loading factor} \\
$N=500$ & 0.5472 & 0.5478 & 0.10 & 0.0392 & 0.0392 & -0.05 \\
$N=2000$ & 0.5472 & 0.5478 & 0.11 & 0.0196 & 0.0201 & 2.57 \\
$N=10000$ & 0.5472 & 0.5474 & 0.04 & 0.0088 & 0.0087 & -0.24 \\ \hline
\multicolumn{7}{l}{Standard deviation principle} \\
$N=500$ & 0.8189 & 0.8185 & -0.05 & 0.1115 & 0.1199 & 7.54 \\
$N=2000$ & 0.8189 & 0.8187 & -0.03 & 0.0577 & 0.0594 & 2.98 \\
$N=10000$ & 0.8189 & 0.8191 & 0.02 & 0.0263 & 0.0262 & -0.22 \\ \hline
\multicolumn{7}{l}{Sharpe ratio principle} \\
$N=500$ & 0.3218 & 0.3259 & 1.29 & 0.0442 & 0.0468 & 5.85 \\
$N=2000$ & 0.3218 & 0.3233 & 0.46 & 0.0229 & 0.0235 & 2.64 \\
$N=10000$ & 0.3218 & 0.3220 & 0.07 & 0.0104 & 0.0105 & 1.73\\
\hhline{=======}
\end{tabular}
\end{table}

Utilizing the results from Theorems~\ref{thm:2}, \ref{thm:4} and \ref{thm:6}, we compute the theoretical standard error of the optimal retention estimators, represented as, for example, $N^{-1/2}\hat c_0^{-1}\sqrt{(\hat c_1, \hat c_2)\hat\Sigma_0(\hat c_1, \hat c_2)^{\tau}}$ under the decreasing loading factor, and compare it with the sample standard deviation of $\{\hat{d}^{*(m)}_{N,\rho_N}\}_{m=1,\ldots,M}$ to assess the validity of the theoretical results. Note that for the computation of standard errors under Theorems \ref{thm:4} and \ref{thm:6}, we utilize the kernel density estimator $\hat{f}_{X_1}(d)$ with a Gaussian kernel function and a bandwidth of 0.1. While alternative kernel functions and bandwidths are possible, we have observed that they exert negligible influence on the computed standard errors, hence we do not delve into further details regarding these alternatives.  
Table~\ref{tab:sim2} summarizes the findings across various $N$ and loading factor rules. Our observations indicate minimal estimation biases of the nonparametrically estimated optimal retention in all scenarios, and in turn, we empirically confirm the consistency of the proposed nonparametric estimators. Moreover, the empirical standard deviations of our estimators closely align with the theoretical standard errors across all cases, which provides empirical validation to the asymptotic properties outlined in Theorems~\ref{thm:2}, \ref{thm:4} and \ref{thm:6}. Additionally, as $N$ becomes large, we note a decline in the relative bias of the estimated optimal retention, as well as the relative difference between the theoretical and empirical standard errors,  consistent with the asymptotic theories.

\section{Real Data analysis} \label{sec:data}

\begin{figure}[!h]
\begin{center}
\begin{subfigure}[h]{0.49\linewidth}
\includegraphics[width=\linewidth]{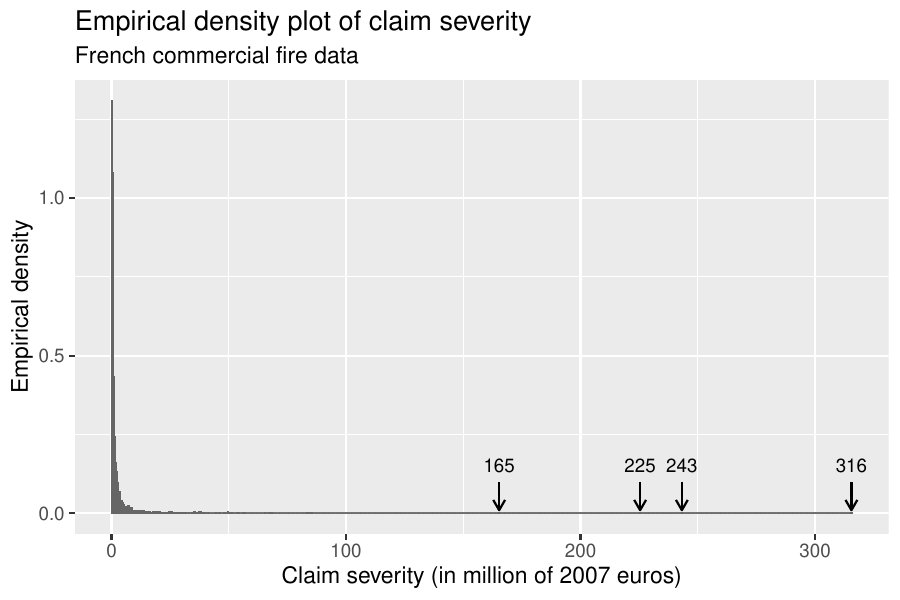}
\end{subfigure}
\begin{subfigure}[h]{0.49\linewidth}
\includegraphics[width=\linewidth]{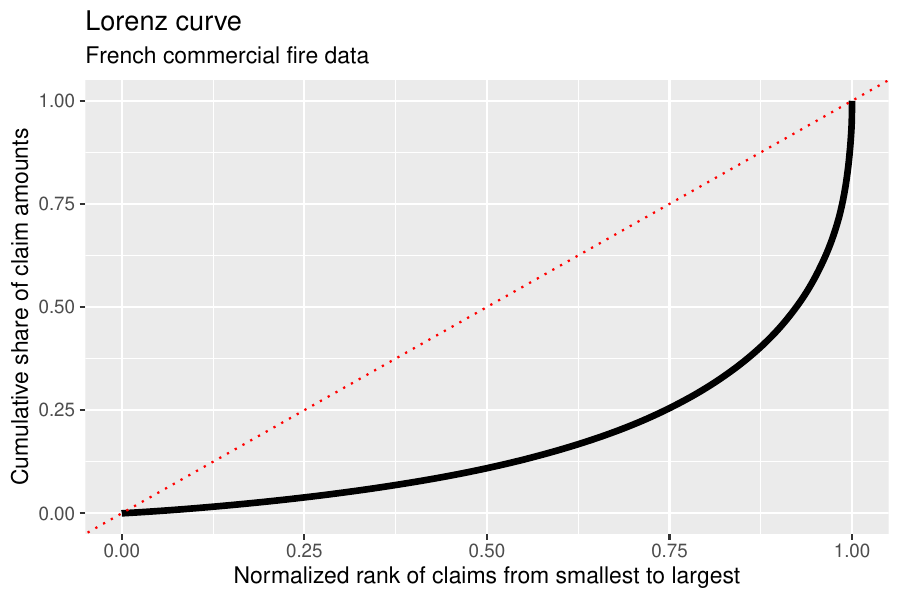}
\end{subfigure}
\end{center}
\vspace{-1.5em}
\caption{\emph{Left panel}: Empirical density plot of claim severity; \emph{Right panel}: Lorenz curve (thick solid curve) of claim severity with the equality line (dotted 45-degree line).}
\label{fig:real_emp}
\end{figure}

We analyze the \texttt{frecomfire} dataset, which consists of 9,613 commercial fire losses located in France, spanning from 1982 to 1996. This dataset is publicly accessible via the \texttt{R} package \texttt{CASdatasets}. The left panel of Figure~\ref{fig:real_emp} displays the empirical density of claim severities, with each claim expressed in million euros (at the 2007 value). The distribution of claim sizes exhibits significant right-skewness, as evidenced by several extreme losses indicated by arrows. In the right panel of Figure~\ref{fig:real_emp}, the Lorenz curve illustrates the cumulative share of claim amounts against the cumulative normalized rank of claims. A substantial deviation of the Lorenz curve from the equality line indicates considerable disparities between large and small claims. The pronounced gap between the Lorenz curve and the equality line reflects the wide dispersion of claim amounts. Notably, the median, mean, and maximum loss amounts are 0.7633, 1.9811, and 315.54, respectively, with the 20 largest loss comprising more than 10\% of the total loss. The heavy-tailed nature of the claim distribution, coupled with several exceptionally large losses, highlights the importance for insurance companies to transfer individual losses, rather than aggregate liabilities, to reinsurers. This motivates the analysis of EoL reinsurance, rather than SL reinsurance, as explored, for instance, by Cai and Tan (2007).

Our primary objective is to investigate the variations in nonparametric estimates of the approximately optimal retention across various effective loading factors $\rho$ and risk levels $p$ under three loading factor rules: decreasing loading factor, standard deviation principle, and Sharpe ratio principle. It is important to highlight that we assess the effective loading factors $\rho$ by using expressions such as \eqref{rho_sd} under the standard deviation principle, rather than relying on the nominal loading factors like $\rho_0$ in \eqref{rho_sd} so that we ensure equitable comparisons among the three loading factor rules.

\begin{figure}[!h]
\begin{center}
\begin{subfigure}[h]{0.49\linewidth}
\includegraphics[width=\linewidth]{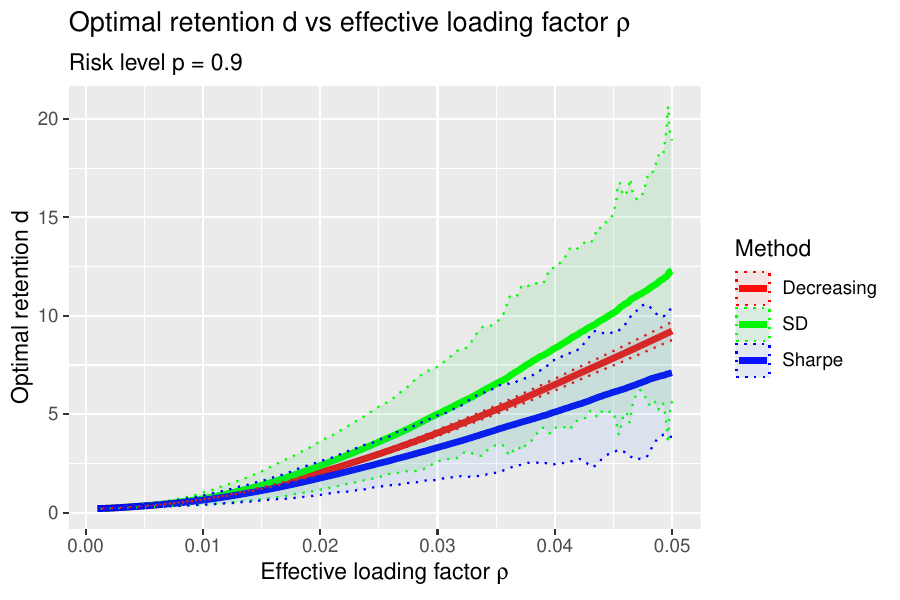}
\end{subfigure}
\begin{subfigure}[h]{0.49\linewidth}
\includegraphics[width=\linewidth]{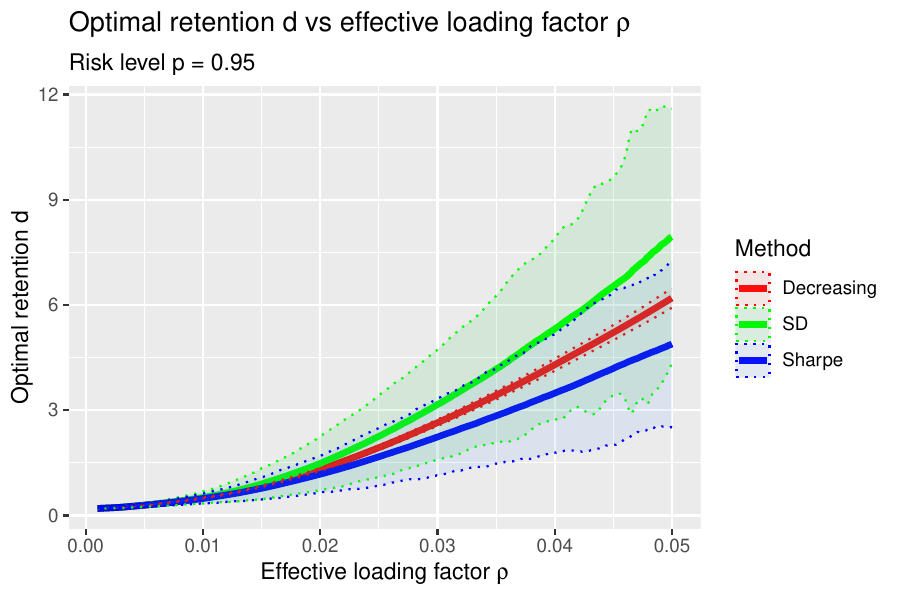}
\end{subfigure}
\end{center}
\vspace{-1.5em}
\caption{Optimal retention versus effective loading factor (solid curves) with fixed $p=0.9$ (left panel) or $p=0.95$ (right panel). The 95\% confidence intervals are displayed as shaded areas.}
\label{fig:real_rho}
\end{figure}

Figure~\ref{fig:real_rho} displays the optimal retention as a function of the effective loading factor $\rho$, with fixed values of $p=0.9$ (left panel) or $p=0.95$ (right panel) under each of the three loading factor rules, accompanied by corresponding 95\% confidence intervals determined based on Theorems~\ref{thm:2}, \ref{thm:4} and \ref{thm:6}. Across all loading factor rules, it is evident that the optimal retention level increases with $\rho$ for any fixed $p$. This observation is intuitive, as a higher $\rho$ implies a greater cost for risk transfer, thereby incentivizing insurers to retain losses up to a higher level. Furthermore, it is observed that the standard deviation loading factor principle yields the highest optimal retention for any fixed $p$ and $\rho$, followed by the decreasing loading factor, and finally the Sharpe ratio principle. This trend can be rationalized by considering that the standard deviation of the excess loss $(X_1-d)_+$ decreases as $d$ increases. Consequently, the effective loading factor under the standard deviation principle diminishes with increasing $d$, encouraging insurers to select a higher retention level to mitigate reinsurance costs. Additionally, the confidence bands under the standard deviation and Sharpe ratio principles are notably wider than those under the decreasing loading factor. This discrepancy arises because the loading factor $\rho$ under either principle, which is contingent on the second moment of the excess loss, may be heavily influenced by extreme losses, leading to increased standard errors. Conversely, the normal-approximated VaR of the total cost relies solely on the excess loss up to its first moment under the decreasing loading factor, resulting in decreased sensitivity of the estimated optimal retention to extreme losses.

\begin{figure}[!h]
\begin{center}
\begin{subfigure}[h]{0.49\linewidth}
\includegraphics[width=\linewidth]{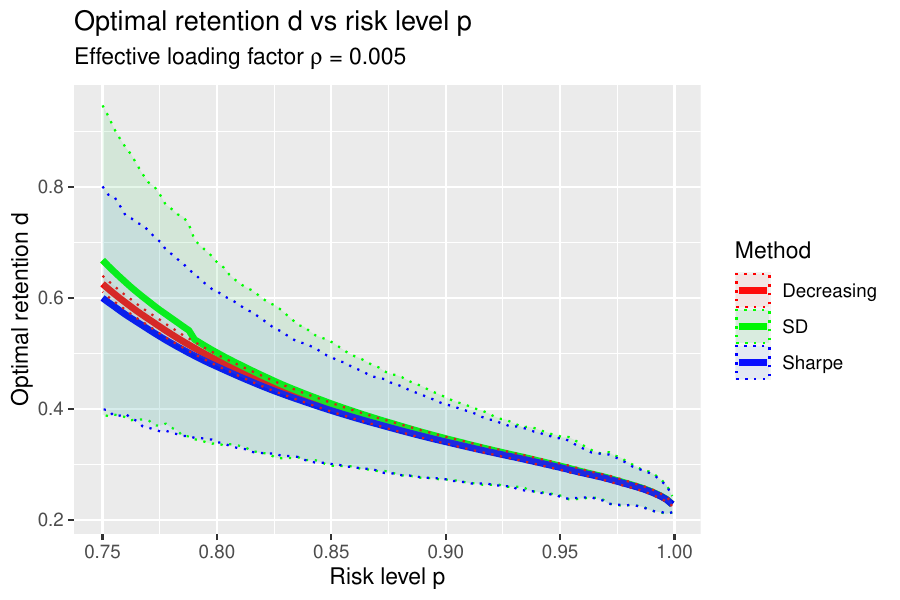}
\end{subfigure}
\begin{subfigure}[h]{0.49\linewidth}
\includegraphics[width=\linewidth]{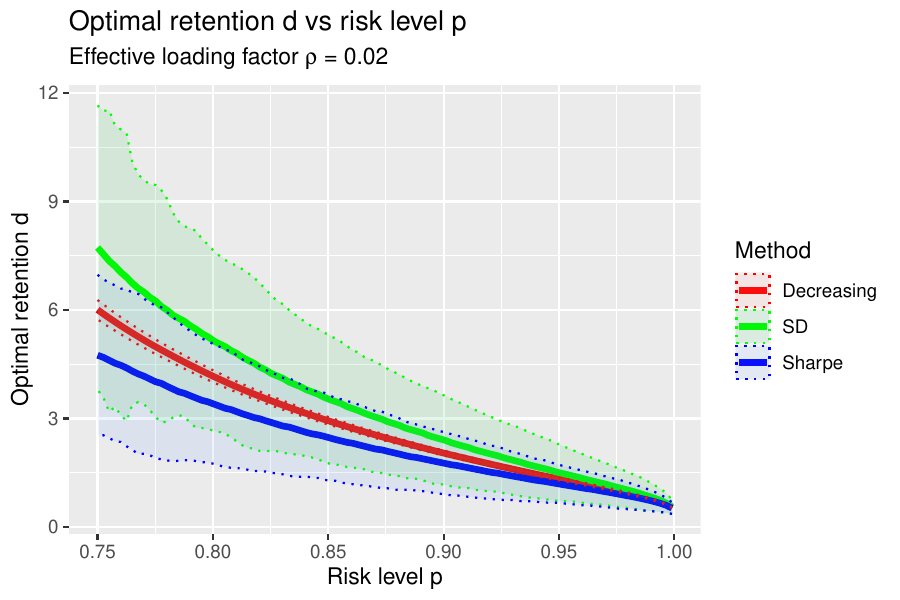}
\end{subfigure}
\end{center}
\vspace{-1.5em}
\caption{Optimal retention versus risk level $p$ (solid curves) with fixed $\rho=0.005$ (left panel) or $\rho=0.02$ (right panel). The 95\% confidence intervals are displayed as shaded areas.}
\label{fig:real_p}
\end{figure}

Figure \ref{fig:real_p} illustrates the optimal retention versus the risk level $p$, with fixed values of $\rho=0.005$ (left panel) or $\rho=0.02$ (right panel) under the three loading factor rules, accompanied by 95\% confidence intervals. Across all loading factor rules, it is observed that the optimal retention decreases as $p$ increases. This outcome is logical, as a higher risk level $p$ signifies insurers' greater aversion to risk, thereby reducing their inclination to retain extreme losses by opting for a smaller retention level. Notably, our proposed method addresses the counterintuitive finding of Cai and Tan (2007) that the optimal retention remains unchanged as $p$ varies.

\section*{References}
\begin{singlespace}
\begin{description}
\item Asimit, A.V., Badescu, A.M. and Cheung, K.C. (2013). Optimal reinsurance in the presence of counterparty default risk. {\it Insurance: Mathematics and Economics, 53(3),690--697.}
\item Asimit, A.V., Badescu, A.M., Haberman, S. and Kim, E.-.S. (2016). Efficient risk allocation within a non-life insurance group under Solvency II Regime. \textit{Insurance: Mathematics and Economics, 66, 69--76.}
\item Asimit, A.V., Badescu, A.M. and Verdonck, T. (2013). Optimal risk transfer under quantile-based risk measures. \textit Insurance: Mathematics and Economics, 53(1), 252--265.
\item Asimit, A.V., Bignozzi, V., Cheung, K.C., Hu, J. and Kim, E.-.S. (2017). Robust and Pareto optimality of insurance contracts. \textit{European Journal of Operational Research, 262(2), 720--732.} 
\item Asimit, A.V., Chi, Y. and Hu, J. (2015). Optimal non-life reinsurance under Solvency II Regime. \textit{Insurance: Mathematics and Economics, 65, 227--237.}
\item Asimit, A.V., Hu, J. and Xie, Y. (2019). Optimal robust insurance with a finite uncertainty set. \textit{Insurance: Mathematics and Economics, 87, 67--81.}
\item Balb\'as, A., Balb\'as, B. and Heras, A. (2009). Optimal reinsurance with general risk measures. {\it Insurance: Mathematics and Economics 44, 374--384.}
\item Balb\'as, A., Balb\'as, B. and Heras, A. (2011). Stable solutions for optimal reinsurance problems involving risk measures. {\it European Journal of Operational Research, 214(3), 796--804.}
\item B\"{a}uerle, N., and Glauner, A. (2018). Optimal risk allocation in reinsurance networks. \textit{Insurance: Mathematics and Economics, 82, 37-47.}
\item Bernard, C. and Ludkovski, M. (2012). Impact of counterparty risk on the reinsurance
market. \textit{North American Actuarial Journal 16(1), 87--111.}
\item Bernard, C., and Tian, W. (2009). Optimal reinsurance arrangements under tail risk measures. {\it Journal of Risk and Insurance 76, 709--725.}
\item Boonen, T. J., and Ghossoub, M. (2023). Bowley vs. Pareto optima in reinsurance contracting. {\it European Journal of Operational Research, 307(1), 382-391.}
\item Boonen, T. J. and Jiang, W. (2022). A marginal indemnity function approach to optimal reinsurance under the Vajda condition. {\it European Journal of Operational Research, 303(2), 928--944.}
\item Boonen, T. J. and Jiang, W. (2024). Robust insurance design with distortion risk measures. {\it European Journal of Operational Research, Forthcoming.}
\item Cai, J., Lemieux, C., and Liu F. (2014) Optimal reinsurance with regulatory initial capital and default risk. \textit{Insurance: Mathematics and Economics 57, 13--24.}
\item Cai, J., Li, J. Y. M. and Mao, T. (2023). Distributionally robust optimization under distorted expectations. {\it Operations Research, Forthcoming.}
\item  Cai, J. and Tan, K.S. (2007). Optimal retention for a stop-loss reinsurance under the VaR and CTE risk measures. {\it Astin Bulletin 37, 93--112.}
\item Cai, J., Tan, K. S., Weng, C. and Zhang, Y. (2008). Optimal reinsurance under VaR and CTE risk measures. {\it Insurance: Mathematics and Economics 43, 185--196.}
\item Cai, J., and Weng C. (2016). Optimal reinsurance with expectile. \textit{Scandinavian Actuarial Journal (7), 624--645.}
\item Chen, Y. (2024). Optimal insurance with counterparty and additive background risk. \textit{ASTIN Bulletin: The Journal of the IAA, 1-22.}
\item Chi, Y. and Tan, K. S. (2021). Optimal incentive-compatible insurance with background risk. \textit{ASTIN Bulletin: The Journal of the IAA, 51(2), 661--688.}
\item Chi, Y. and Tan, K. S. (2013). Optimal reinsurance with general premium principles. {\it Insurance: Mathematics and Economics 52, 180--189.}
\item Denneberg, D. (1994a). Non-additive measure and integral. \textit{Kluwer Academic Publishers, Dordrecht.}
\item Denneberg, D. (1994b). Conditioning (updating) non-additive measures. \textit{Annals of Operations Research, 52, 21--42.}
\item Embrechts, P. and Hofert, M. (2013). A note on generalized inverses, \textit{Mathematical Methods of Operations Research, 77, 423--432.}
\item F\"{o}llmer, H. and Schied, A. (2011). 
Stochastic Finance: An Introduction in Discrete Time, \textit{Third ed., Walter de Gruyter}.
\item Gollier, C. (2014). Optimal insurance design of ambiguous risks. \textit{Economic Theory 57, 555--576.}
\item Kaluszka, M. (2001). Optimal reinsurance under mean-variance premium principles. {\it Insurance: Mathematics and Economics 28, 61--67.}
\item R\"{u}schendorf, L. (2013). \textit{Mathematical Risk Analysis: Dependence, Risk Bounds, Optimal Allocations and Portfolios. Springer.}
\item Wang, R., Wei, Y. and Willmot, G. E. (2020). Characterization, robustness and aggregation of signed Choquet integrals. \textit{Mathematics of Operations Research, 45(3), 993--1015.}
\item Yaari, M. E. (1987). The dual theory of choice under risk. \textit{Econometrica: Journal of the Econometric Society, 95--115.}
  \end{description}
\end{singlespace}

\newpage
\begin{center}
\begin{LARGE}
{\bf Supplementary materials for ``A Revisit of the Optimal Excess-of-Loss Contract"}
\end{LARGE}
\end{center}

\setcounter{section}{0}

\section{Numerical study: Comparing EoL and SL approaches}\label{supp:sec:num}
In this section, we substantiate the assertions presented in Section \ref{sec:setup:bg} of the manuscript regarding the comparison between the EoL and SL approaches. This is accomplished through a numerical investigation with a small $N$ supported by theoretical reasoning. We simulate $X_i$ from a Pareto (type-II) distribution with pdf $f_X(x)=(\alpha/\lambda)(1+x/\lambda)^{-(\alpha+1)}$ with shape parameter $\alpha=9$ and scale parameter $\lambda=8$, such that the mean is $E[X_i]=\alpha/(\lambda-1)=1$. We choose $p=0.75$ for the risk level, $\rho=0.2$ for the loading factor, and consider $N=2,3,5,10$. We first compute the true VaR of the total cost $\text{VaR}_p\big(T(d,N,\rho)\big)$ using \eqref{T_2_EVP}, where $\text{VaR}_p(\sum_{i=1}^{N}(X_i\wedge d))$ and $\mathbb{E}\{(X_1-d)_{+}\}$ are approximated, respectively, by the empirical $p$-VaR and expectation of $\sum_{i=1}^{N}(X_i\wedge d)$ and $(X_1-d)_{+}$ from $B=50,000$ simulated samples. For example, we compute $\mathbb{E}\{\sum_{i=1}^{N}(X_i\wedge d)\}\approx (1/B)\sum_{j=1}^{B}\sum_{i=1}^{N}(X_{ij}\wedge d)$ with $X_{ij}$ iid sampled from the Pareto distribution for $i=1,\ldots,N$ and $j=1,\ldots,B$.

\begin{figure}[ht]
\begin{center}
\begin{subfigure}[h]{0.49\linewidth}
\includegraphics[width=\linewidth]{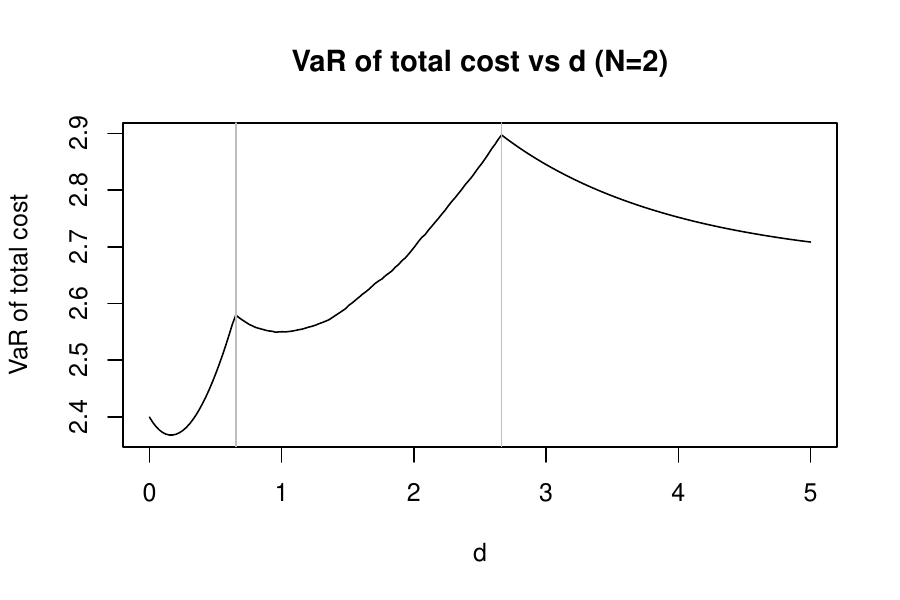}
\end{subfigure}
\begin{subfigure}[h]{0.49\linewidth}
\includegraphics[width=\linewidth]{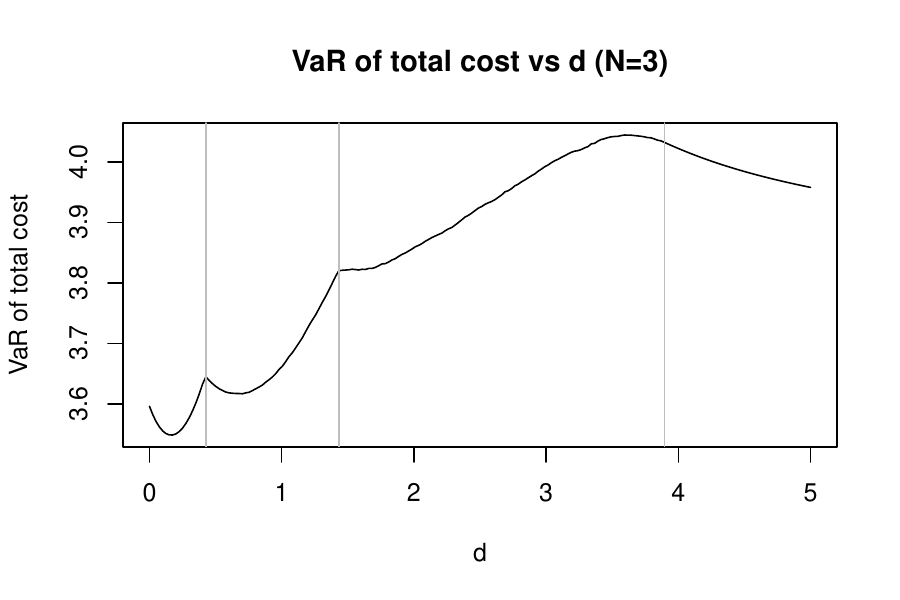}
\end{subfigure}
\begin{subfigure}[h]{0.49\linewidth}
\includegraphics[width=\linewidth]{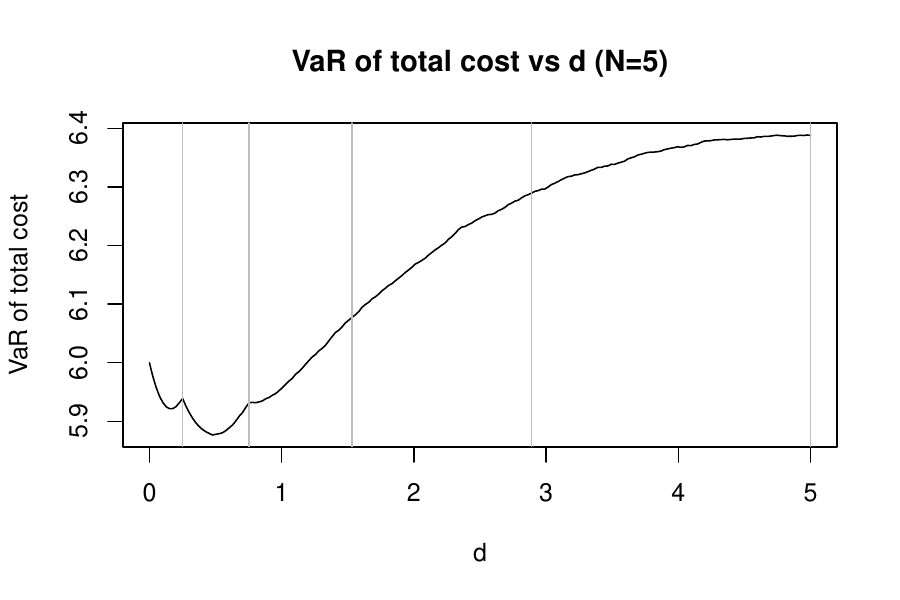}
\end{subfigure}
\begin{subfigure}[h]{0.49\linewidth}
\includegraphics[width=\linewidth]{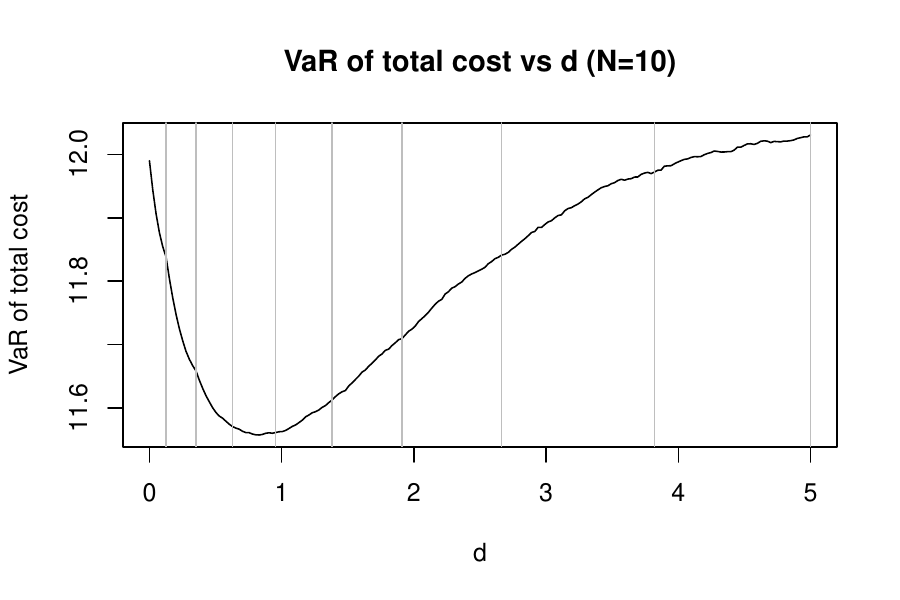}
\end{subfigure}
\end{center}
\vspace{-1.5em}
\caption{$\text{VaR}_p\big(T(d,N,\rho)\big)$ versus $d\in[0,5]$ for various $N$. Gray vertical lines represent the turning points of the curves.}
\label{fig:num1}
\end{figure}

Figure~\ref{fig:num1} plots $\text{VaR}_p\big(T(d,N,\rho)\big)$ as a function of $d$ for various $N$. We note that $\text{VaR}_p\big(T(d,N,\rho)\big)$ is piecewise differentiable except for $N$ turning points, which are illustrated by the gray dotted vertical lines added in Figure~\ref{fig:num1}.  In addition, if we denote the $i$-th turning point as $\tilde{d}_{N}(i)$ for $i=1,\ldots,N-1$, we observe that $\tilde{d}_{N}(i)=\{d:P(\sum_{i=1}^N(X_i\wedge d)\geq(N-i+1)d)=1-p\}$; indeed, the density function of $\sum_{i=1}^N(X_i\wedge d)$ exhibits jump points at integer multiples of $d$, causing the turning points.

We then numerically calculate $d^*$, i.e., minimize $\text{VaR}_p\big(T(d,N,\rho)\big)$ over $d$, and compute the probability $P(T(d^*,N,\rho)>\text{VaR}_p(T(d^*,N,\rho)))$ for various $N$. An interesting series of observations is further noted.

\emph{First,} the calculated value of $d^*$ is 0.1633, 0.1643, 0.5025, 0.8779, respectively, for $N=2,3,5,10$. This coincides with the analysis in Theorem~\ref{thm:1} that the optimal retention increases with $N$ if $\rho$ is fixed.

\emph{Second,} the calculated probability $P(T(d^*,N,\rho)>\text{VaR}_p(T(d^*,N,\rho)))$ is 0, 0, 0.25 and 0.25, respectively, for $N=2,3,5,10$. As $N$ is large enough, the probability will be exactly 0.25, and otherwise, the probability will be exactly zero.

\emph{Third,} we observe from the figure that $P(T(d^*,N,\rho)>\text{VaR}_p(T(d^*,N,\rho)))=0$ if $d^*\leq\tilde{d}_{N}(1)$ and $P(T(d^*,N,\rho)>\text{VaR}_p(T(d^*,N,\rho)))=0.25$ if $d^*>\tilde{d}_{N}(1)$. Indeed, one can justify it theoretically as follows. From the definition of $\tilde{d}_{N}(i)$, if $d^*\leq\tilde{d}_{N}(1)$, we have
\begin{align*}
1-p&=P(\sum_{i=1}^NX_i\wedge \tilde{d}_{N}(i)\geq N\tilde{d}_{N}(i))=P(X_i\geq \tilde{d}_{N}(i))^N\\
&\leq P(X_i\geq d^*)^N = P(\sum_{i=1}^NX_i\wedge d^*\geq Nd^*).
\end{align*}
Since $P(\sum_{i=1}^NX_i\wedge d^*\geq Nd^*)\geq 1-p$, $\sum_{i=1}^NX_i\wedge d^*$ is upper bounded by $Nd^*$ and hence $P(\sum_{i=1}^NX_i\wedge d^*> Nd^*)=0$, we have $\text{VaR}_p(\sum_{i=1}^NX_i\wedge d^*)=Nd^*$ and hence $P(\sum_{i=1}^NX_i\wedge d^*> \text{VaR}_p(\sum_{i=1}^NX_i\wedge d^*))=0$. If $d^*>\tilde{d}_{N}(1)$, we have $\text{VaR}_p(\sum_{i=1}^NX_i\wedge d^*)<Nd^*$, and the distribution function of $\sum_{i=1}^NX_i\wedge d^*$ is continuous on $(0,Nd^*)$. Hence, $P(\sum_{i=1}^NX_i\wedge d^*> \text{VaR}_p(\sum_{i=1}^NX_i\wedge d^*))=1-p$ by the basic definition of quantile. Therefore, we conclude that the VaR of the total cost with the optimal retention is appropriate if and only if the optimal retention is above the first turning point.

\emph{Fourth,} one can also show that $P(T(d^*,N,\rho)>\text{VaR}_p(T(d^*,N,\rho)))=1-p$ if and only if $P(X_i\geq d^*)\leq (1-p)^{1/N}$. With a larger $N$, this condition is more likely to hold. With $N=1$, i.e., the SL approach following Cai and Tan (2007), we have $d^*=F_{X_1}^{-}(1-1/(1+\rho))$ given $1-p<(1+\rho)^{-1}$, and hence $P(X_i\geq d^*)=(1+\rho)^{-1}>(1-p)$, meaning that the condition never holds.

Overall, while $\p\left(T(d^*,1,\rho)>\VaR_p(T(d^*,1,\rho))\right)=0$ under the SL approach, we empirically and theoretically show that $\p\left(T(d^*,N,\rho)>\VaR_p(T(d^*,N,\rho))\right)=1-p$, the correct level, for sufficiently large $N$ under the EoL approach. Hence, the EoL optimal retention would not inherit the same counter-intuitive property as the SL optimal retention.

\section{Approximating VaR of total cost by Edgeworth expansion} \label{apx:sec:edgeworth}
To address the issue of insufficient accuracy outlined by Section \ref{sec:sim1} under the constant loading factor rule, we employ Edgeworth expansion to improve the approximation precision from \eqref{appr} in the manuscript to obtain $G^{(2)}_{N,\rho}$ and $G^{(3)}_{N,\rho}$. Suppose that $Z_1,\ldots,Z_N$ are iid random variables with zero mean, unit variance, and $E[Z_1^4]<\infty$. Then, standard Edgeworth expansion yields
\begin{align}\label{eq:edgeworth}
P\left(\frac{1}{\sqrt{N}}\sum_{i=1}^NZ_i\leq x\right)
&=\Phi(x)-\frac{1}{\sqrt{N}}p_1(x)\phi(x)+\frac{1}{N}p_2(x)\phi(x)+o(N^{-1}),
\end{align}
where $p_1(x)=-\kappa_3H_2(x)/6$, $p_2(x)=-(\kappa_4H_3(x)/24+\kappa_3^2H_5(x)/72)$. Here, $\kappa_3=E[(Z_1-E[Z_1])^3]$ and $\kappa_4=E[(Z_1-E[Z_1])^4]-3E[(Z_1-E[Z_1])^2]^2$ are the moment quantities, and $H_2(x)=x^2-1$, $H_3(x)=x^3-3x$ and $H_5(x)=x^5-10x^3+15x$ are the Hermite polynomials. From \eqref{eq:edgeworth} and \eqref{T_2_EVP}, one can apply the Cornish-Fisher expansion and write
\begin{align}\label{eq:edgeworth:VaR}
\text{VaR}_p\left(T(d,N,\rho)\right)&=\sqrt{N}\sqrt{\mu_2(d)-\mu_1^2(d)}\left[\Phi^-(p)+\frac{1}{\sqrt{N}}\tilde{p}_1(p;d)+\frac{1}{N}\tilde{p}_2(p;d)\right]\\
&\hspace{1cm}+N\E[X_1]+N\rho\nu_1(d)+o\left(\frac{1}{\sqrt{N}}\right),\nonumber
\end{align}
where $\tilde{p}_1(p;d)=-\frac{\tilde{\kappa}_3(d)}{6}H_2(p)$ and
\begin{equation*}
\tilde{p}_2(p;d)=\frac{\tilde{\kappa}_4(d)}{24}H_3(p)+\frac{\tilde{\kappa}_3(d)^2}{72}\left(H_5(p)+2H'_2(p)H_2(p)-pH_2(p)^2\right)
\end{equation*}
with
\begin{equation*}
\tilde{\kappa}_3(d)=\frac{\E[(X_1\wedge d-\E[X_1\wedge d])^3]}{\E[(X_1\wedge d-\E[X_1\wedge d])^2]^{3/2}},\qquad \tilde{\kappa}_4(d)=\frac{\E[(X_1\wedge d-\E[X_1\wedge d])^4]}{\E[(X_1\wedge d-\E[X_1\wedge d])^2]^{2}}-3.
\end{equation*}
We alternatively propose to compute the approximate optimal retention by minimizing
\begin{align}\label{eq:G2}
G_{N,\rho}^{(2)}(d)=N\E[X_1]+N\rho\nu_1(d)+\sqrt{N}\sqrt{\mu_2(d)-\mu_1^2(d)}\left[\Phi^-(p)+\frac{1}{\sqrt{N}}\tilde{p}_1(p;d)\right]
\end{align}
or 
\begin{eqnarray}\label{eq:G3}
&&G_{N,\rho}^{(3)}(d)=N\E[X_1]+N\rho\nu_1(d)\\
&&\hspace{0.75in}+\sqrt{N}\sqrt{\mu_2(d)-\mu_1^2(d)}\left[\Phi^-(p)+\frac{1}{\sqrt{N}}\tilde{p}_1(p;d)+\frac{1}{N}\tilde{p}_2(p;d)\right],\nonumber
\end{eqnarray}
where the error terms underlying $G_{N,\rho}^{(2)}$ and $G_{N,\rho}^{(3)}$ are, respectively, $o(1)$ and $o(N^{-1/2})$.

\end{document}